\documentclass[%
 preprint, 
 amsmath,amssymb,
 aps, pre, physrev, showpacs, longbibliography, tightenlines
]{revtex4-2}

\usepackage{graphicx}
\usepackage{dcolumn}
\usepackage{bm}

\usepackage{enumerate}
\usepackage{subfigure}

\allowdisplaybreaks

\begin{document}

\preprint{APS/123-QED}

\title{\textbf{Radiometer effect on an infinitely thin circular disk} 
}%

\author{Takuma Tomita}
\author{Satoshi Taguchi}%
 \email{Contact author: taguchi.satoshi.5a@kyoto-u.ac.jp}
\author{Tetsuro Tsuji}
\affiliation{%
  Graduate School of Informatics, Kyoto University, Kyoto 606-8501, Japan
}%



\date{\today}

\begin{abstract}
The radiometer effect is a self-thermophoretic phenomenon in which a thin body with a temperature difference between its two sides experiences a force in a rarefied gas. We numerically investigate this effect for an infinitely thin circular disk freely translating at its terminal velocity in an otherwise quiescent gas. The problem is formulated using the linearized Bhatnagar--Gross--Krook model of the Boltzmann equation with the diffuse reflection boundary condition. Owing to linearity, the problem is decomposed into fundamental auxiliary problems, allowing the radiometric force and the drag force to be evaluated separately and the terminal velocity to be determined from their balance. A major difficulty arises from the sharp disk edge, which generates discontinuities in the velocity distribution function that propagate into the gas. Accurate resolution of these discontinuities is essential for capturing the edge-localized stress associated with the radiometric force in the near-continuum regime. The discontinuities are explicitly accounted for using a characteristic-based numerical scheme, together with the appropriate far-field asymptotics for the unbounded domain. The radiometric force, the terminal velocity, and the flow and stress fields are obtained over a wide range of Knudsen numbers. The dimensionless radiometric force increases monotonically with the Knudsen number, scaling as $\mathrm{Kn}^{3/2}$ in the continuum limit and approaching a constant in the free-molecular limit. The terminal velocity exhibits a nonmonotonic dependence on the Knudsen number and slightly exceeds its free-molecular-limit value at intermediate Knudsen numbers. In the continuum limit, the magnitude of the edge-localized variation of the normal-stress difference appears to remain finite, whereas its characteristic width is of the order of the molecular mean free path. The resulting numerical data also provide reference solutions for future studies of sharp-edge kinetic problems. Finally, the computed radiometric force is independently validated using the symmetric relation for the linearized Boltzmann equation.
\end{abstract}

\maketitle

\newpage

\section{\label{sec:introduction}Introduction}

The radiometer effect, classically exemplified by the well-known Crookes radiometer \cite{PTotRSoL_Crookes_1874}, is a manifestation of self-thermophoresis in a rarefied gas \cite{V_KetsdeverEtAl_2012,JoMS_WadsworthMuntz_1996,MaCiS_OtaEtAl_2001,PRE_SeldenEtAl_2009,JoFM_SeldenEtAl_2009,CMaMP_Anikin_2011,PoF_ChenEtAl_2012,PRE_TaguchiAoki_2015,JoCP_DechristeMieussens_2016}. A thin vane with a temperature difference between its two sides experiences a force, called the radiometric force, and translates with the hotter side trailing [Fig.~\ref{fig:problem}(a)]. 
The force originates from the asymmetry of momentum exchange between the hotter and colder sides: the hotter side reflects gas molecules at higher speed than the colder side on average, resulting in a recoil force exerted by the gas on the vane.
The radiometer effect has recently attracted renewed interest for its potential applications to microfluidic devices \cite{JoMS_WadsworthMuntz_1996,MaCiS_OtaEtAl_2001,PRL_PassianEtAl_2003,PRL_Wuerger_2011,APL_StrongrichAlexeenko_2015,JoMS_StrongrichEtAl_2017,PRF_BaierEtAl_2017,JoFM_LotfianRoohi_2018,M&N_WangEtAl_2020}, aerosol engineering \cite{PotNAoS_Keith_2010, Smalley_2018}, and near-space flight \cite{AM_CortesEtAl_2020, N_SchaferEtAl_2025}.

One important feature of the radiometer effect is that the spatial distribution of the force strongly depends on the degree of gas rarefaction, characterized by the Knudsen number $\mathrm{Kn}$, defined by the ratio of the molecular mean free path to the characteristic dimension of the vane. For large $\mathrm{Kn}$ (i.e., the free-molecular regime), the radiometric force is distributed over the entire surface of the vane. In contrast,
for small $\mathrm{Kn}$ (i.e., the near-continuum regime), the force becomes localized near the edge of the vane \cite{PRE_SeldenEtAl_2009,JoFM_SeldenEtAl_2009,JoFM_TaguchiAoki_2012,PRE_TaguchiAoki_2015}, a phenomenon commonly referred to as the edge effect. Consequently, the quantitative behavior of the radiometer effect differs markedly from that of self-thermophoresis of a particle with a smooth surface, such as a sphere.

The principal objective of this paper is to clarify how the radiometric force and the resulting motion of a thin vane vary across the entire range of gas rarefaction. Particular attention is paid to the edge-localized stress associated with the force in the near-continuum regime. As the simplest geometry retaining this essential edge effect, we consider an infinitely thin circular disk freely moving in an infinite domain. The effect of the vane thickness, which is known to reduce the radiometric force to some extent \cite{JoFM_VenturaEtAl_2013}, is not incorporated in the present analysis.

A major difficulty in obtaining quantitatively reliable solutions to this problem arises from the sharp edge of the disk. It is known that a sharp edge generates discontinuities in the velocity distribution function (VDF), which propagate into the gas \cite{JoFM_TaguchiAoki_2012}. In the present three-dimensional geometry, the resulting discontinuity structure is considerably more complicated than in two-dimensional sharp-edge problems. Accurate resolution of these discontinuities is particularly important for describing the edge-localized stress in the near-continuum regime. Even within a model kinetic equation, resolving this structure over a wide range of Knudsen numbers requires substantial numerical effort.

We formulate the problem using the Bhatnagar--Gross--Krook (BGK) model \cite{AF_Welander_1954,PR_BhatnagarEtAl_1954} of the Boltzmann equation with the diffuse reflection boundary condition. The system is linearized under the assumptions of a small temperature difference and slow disk motion. This linearization also allows the problem to be decomposed into fundamental auxiliary problems. The use of the BGK model makes the kinetic computation tractable while allowing us to devote sufficient numerical resolution to the discontinuity structure generated by the disk edge.

For the numerical analysis, we employ the characteristic-based scheme developed in \cite{JoFM_TomitaEtAl_2026}, which is formulated directly from the integral form of the kinetic equation and explicitly accounts for the discontinuities in the VDF. The appropriate far-field asymptotics are also incorporated into the numerical formulation to properly treat the unbounded physical domain. These features enable highly resolved computations of the radiometric force, the terminal velocity, and the flow and stress fields over a wide range of Knudsen numbers. The resulting numerical data can also serve as reference solutions for future studies of sharp-edge kinetic problems. As an independent check of the numerical results, we evaluate the radiometric force using the symmetric relation for the linearized Boltzmann equation \cite{JoSP_Takata_2009,JoSP_Takata_2009a}.

The remainder of the paper is structured as follows. In Sec.~\ref{sec:problem_and_decomposition}, we formulate the problem and decompose it into the three fundamental problems, followed by a short summary of the numerical scheme in Sec.~\ref{sec:numerical_method}. In Sec.~\ref{sec:numerical_results}, we present the numerical results for the radiometric force, the terminal velocity of the disk, the flow field, and the normal-stress distribution on the disk, together with comparisons with available experimental data. Section~\ref{sec:reciprocal_relation} is devoted to numerical validation using the symmetric relation derived in \cite{JoSP_Takata_2009}. In Sec.~\ref{sec:conclusion}, we conclude the paper.

\section{\label{sec:problem_and_decomposition}Formulation}

\subsection{\label{subsec:problem}Problem}

\begin{figure*}
\centering
\subfigure[]{\includegraphics[scale=0.95]{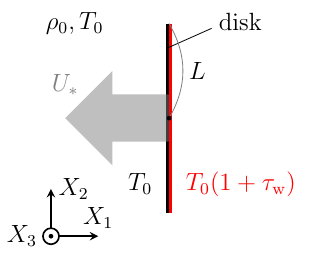}}
\subfigure[]{\includegraphics[scale=0.95]{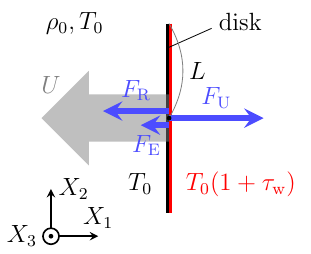}}
\subfigure[]{\includegraphics[scale=0.95]{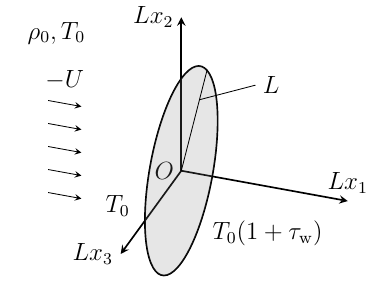}}
\caption{Schematics of the problems. 
(a) Problem A: An infinitely thin circular disk with one side heated undergoes translational motion with the terminal velocity $U_*$.
(b) Problem B: Uniform translational motion of the same disk as in Problem A with a prescribed velocity $U$.
(c) Three-dimensional representation of Problem B in the Cartesian coordinate system moving with the disk.
}
\label{fig:problem}
\end{figure*}

In this paper, we consider the following problem as a simple model of a radiometer vane.

Problem A: Consider an infinitely thin circular disk with radius $L$ that is free to move along its axis of symmetry in an otherwise quiescent rarefied gas [Fig.~\ref{fig:problem}(a)]. Let $X_i$ ($i=1,2,3$) be Cartesian coordinates, with the $X_1$-axis parallel to the axis of symmetry of the disk. Far from the disk, the gas is in an equilibrium state at rest with density $\rho_0$ and temperature $T_0$. One side of the disk, facing the positive $X_1$-direction, is maintained at a higher temperature $T_0(1+\tau_{\mathrm{w}})$ ($\tau_{\mathrm{w}} > 0$), while the opposite side is maintained at the ambient temperature $T_0$. 
Owing to the resulting radiometric force, the disk moves from the heated side toward the unheated side, i.e., in the negative $X_1$-direction. Suppose that the radiometric force is balanced by the drag force, so that the disk reaches a terminal velocity. Determine the terminal velocity of the disk $U_{*}$ and the behavior of the gas around the disk. In the present configuration, $U_{*} < 0$. The basic assumptions of the analysis are as follows:
\begin{enumerate}[(i)]
\item The behavior of the gas is described by the BGK model of the Boltzmann equation.
\item Gas molecules undergo diffuse reflection at the disk surface.
\item The temperature perturbation is small, i.e., $\tau_{\mathrm{w}} \ll 1$.
\item The terminal velocity is much smaller than the thermal speed,
$|U_{*}| \ll (2RT_0)^{1/2}$,
where $R$ is the specific gas constant.
\end{enumerate}
Assumptions (iii) and (iv) allow the governing equation and the boundary conditions to be linearized around the reference equilibrium state at rest with density $\rho_0$ and temperature $T_0$.

The difficulty is that the terminal velocity $U_{*}$ is not known a priori, since it is determined as a response to the temperature perturbation $\tau_{\mathrm{w}}$. 
To overcome this difficulty, we introduce a second problem, in which the velocity of the disk is prescribed as a control parameter.

Problem B: Consider an infinitely thin circular disk of radius $L$ moving uniformly in the $X_1$-direction with a prescribed velocity $U$, which is not necessarily equal to $U_*$ [Fig.~\ref{fig:problem}(b)]. The geometry, the far-field equilibrium condition, and the temperature distribution on the disk are the same as those in Problem A. Investigate the behavior of the gas under assumptions (i)--(iii) and the additional assumption
\begin{enumerate}[(iv)']
\item The prescribed disk velocity satisfies $|U| \ll (2RT_0)^{1/2}$.
\end{enumerate}
In general, the drag and radiometric forces do not balance each other. Therefore, to maintain the prescribed uniform motion, an external force must be applied to the disk. Let $F_{\mathrm{U}}$, $F_{\mathrm{R}}$, and $F_{\mathrm{E}}$ denote the drag, radiometric, and external forces, respectively.
They act normal to the disk. Then,
\begin{align}
\label{eq:balance_in_ProbB}
    F_{\mathrm{U}} + F_{\mathrm{R}} + F_{\mathrm{E}} = 0,
\end{align}
(see Fig.~\ref{fig:problem}(b)). If the prescribed velocity $U$ is adjusted so that the external force $F_{\mathrm{E}}$, or equivalently the net force exerted by the gas, $F_{\mathrm{U}} + F_{\mathrm{R}}$, vanishes,  Problem B reduces to Problem A, and the corresponding value of $U$ is precisely the terminal velocity $U_*$. This adjustment can be carried out semi-analytically after the drag and radiometric forces have been computed numerically by the decomposition procedure described in Secs.~\ref{subsec:decomposition} and \ref{subsec:drag_force}.
This strategy is analogous to that employed by \cite{PRE_TaguchiAoki_2015} for a nonlinear problem. In the present linear setting, however, the decomposition enables the drag and radiometric forces to be evaluated independently, allowing the terminal velocity to be determined in a semi-analytical manner.

\subsection{\label{subsec:formulation}Basic equations}

We formulate only Problem B, since Problem A is recovered as a special case, as explained above.
A reference frame moving with the disk is convenient, since the uniform translational motion of the disk is equivalent, by a Galilean transformation, to a stationary disk placed in a uniform flow of the same speed in the opposite direction. Consequently, Problem B can be formulated as a steady boundary-value problem.
Accordingly, we introduce Cartesian coordinates $Lx_i$ ($i = 1,2,3$) with the origin fixed at the center of the disk and the positive $x_1$-axis directed toward the heated side [Fig.~\ref{fig:problem}(c)]. The two sides of the disk are denoted by $S^+$ and $S^-$, corresponding to the positive and negative sides of the $x_1$-axis, respectively:
\begin{align}
\label{eq:def_S_pm}
    S^{\pm} = \left\{ (x_1, x_2, x_3) \mid x_1 = 0_{\pm}, \,\, x_2^2 + x_3^2 \leq 1 \right\}.
\end{align}
Throughout this paper, $0_{\pm}$ denotes the limit $\lim_{\varepsilon \to \pm 0} \varepsilon$.
In the present coordinate system, $S^{+}$ and $S^{-}$ are maintained at the uniform temperatures $T_{\mathrm{w}}^{+} = T_0(1+\tau_{\mathrm{w}})$ and $T_{\mathrm{w}}^{-} = T_0$, respectively. Far from the disk, the gas is in an equilibrium state with density $\rho_0$, temperature $T_0$, and flow velocity $(-U,0,0)$. 

We next introduce the basic notation. Let $(2RT_0)^{1/2} \zeta_i$ denote the molecular velocity and $\rho_0(2RT_0)^{-3/2} (1 + \phi(\bm{x}, \bm{\zeta})) E$ the velocity distribution function (VDF), where $E = E(\bm{\zeta}) := \pi^{-3/2} \exp(-\zeta_j^2)$. The density, flow velocity, temperature, pressure, stress tensor, and heat flow are denoted by $\rho_0 (1 + \omega(\bm{x}))$, $(2RT_0)^{1/2} u_i(\bm{x})$, $T_0(1 + \tau(\bm{x}))$, $p_0(1+P(\bm{x}))$, $p_0(\delta_{ij} + P_{ij}(\bm{x}))$, and $p_0 (2RT_0)^{1/2} Q_i(\bm{x})$, respectively, where $p_0 = \rho_0 RT_0$ is the pressure in the reference state and $\delta_{ij}$ is the Kronecker delta. These macroscopic quantities are expressed in terms of moments of $\phi$ with respect to the molecular velocity:
\begin{subequations}
\label{eq:def_macro}
\begin{align}
\label{eq:def_out}
    &\omega = \langle \phi \rangle, \quad u_i = \langle \zeta_i \phi \rangle, \quad \tau = \frac{2}{3} \langle (\zeta_j^2 - \frac{3}{2}) \phi \rangle, \\
\label{eq:def_ppq}
    &P = \omega + \tau = \frac{2}{3} \langle \zeta_j^2 \phi \rangle, \quad P_{ij} = 2 \langle \zeta_i \zeta_j \phi \rangle, \quad Q_i = \langle \zeta_i \zeta_j^2 \phi \rangle - \frac{5}{2} u_i.
\end{align}
\end{subequations}
Here, the brackets denote
\begin{align}
\label{eq:def_bracket}
    \langle f(\bm{\zeta}) \rangle = \int_{\mathbb{R}^3} f(\bm{\zeta}) E\, \mathrm{d}\bm{\zeta},
\end{align}
where $\mathrm{d}\bm{\zeta} = \mathrm{d} \zeta_1 \mathrm{d} \zeta_2 \mathrm{d} \zeta_3$.

Problem B is formulated as the following boundary-value problem, consisting of the linearized BGK equation and the diffuse reflection boundary condition on the disk, supplemented by the condition at infinity:
\begin{subequations}
\label{eq:basic_system}
\begin{align}
\label{eq:LBGK}
    &\zeta_i \frac{\partial \phi}{\partial x_i} = \frac{1}{\kappa} \mathcal{L}(\phi), \\
\label{eq:bc_Sp_and_Sm}
    &\text{b.c. } \phi = \begin{cases}
        \sigma_{\mathrm{w}}^{+}(\phi) + (\zeta_j^2 - 2) \tau_{\mathrm{w}} \quad
        &\text{for } \zeta_1 > 0 \quad (\bm{x} \in S^{+}), \\
        \sigma_{\mathrm{w}}^{-}(\phi) \quad
        &\text{for } \zeta_1 < 0 \quad (\bm{x} \in S^{-}), \\
    \end{cases} \\
\label{eq:bc_inf}
    &\text{b.c. } \phi \to 2 u_{\infty} \zeta_1 \quad (|\bm{x}| \to \infty).
\end{align}
\end{subequations}
Here, $\mathcal{L}$ represents the linearized BGK collision operator,
\begin{align}
\label{eq:def_L}
    &\mathcal{L}(\phi) = -\phi + \omega + 2\zeta_i u_i + (\zeta_j^2 - \frac{3}{2}) \tau,
\end{align}
and $\sigma_{\mathrm{w}}^{\pm}$ is defined by
\begin{align}
\label{eq:def_sigmaw}
    \sigma_{\mathrm{w}}^{\pm}(\phi) = 2\sqrt{\pi} \int_{0}^{\mp\infty} \zeta_1 \left( \int_{-\infty}^{\infty} \int_{-\infty}^{\infty} \phi E \mathrm{d}\zeta_2\mathrm{d}\zeta_3 \right) \mathrm{d}\zeta_1.
\end{align}
The parameter $\kappa$ in \eqref{eq:LBGK} is related to the Knudsen number $\mathrm{Kn}$ by 
\begin{align}
\label{eq:def_kappa}
    \kappa = \frac{\sqrt{\pi}}{2} \mathrm{Kn} = \frac{\sqrt{\pi}}{2} \frac{\ell_0}{L},
\end{align}
where $\ell_0$ is the mean free path in the reference equilibrium state. 
In \eqref{eq:bc_inf}, $u_{\infty} := -U (2RT_0)^{-1/2}$ is the dimensionless velocity of the uniform flow at infinity and satisfies $|u_{\infty}| \ll 1$ by assumption (iv)' in Sec.~\ref{subsec:problem}.

\subsection{\label{subsec:decomposition}Decomposition}

Exploiting the linearity of the system \eqref{eq:basic_system}, we separate the contributions from the boundary conditions by superposition. To this end, we decompose the VDF and the macroscopic quantities as
\begin{subequations}
\label{eq:superposition}
\begin{align}
\label{eq:superposition_VDF}
    &\phi = u_{\infty} \phi^{\mathrm{U}} + \frac{\tau_{\mathrm{w}}}{2} \phi^{\mathrm{R}} + \frac{\tau_{\mathrm{w}}}{2} \phi^{\mathrm{T}}, \\
\label{eq:superposition_macro}
    &h = u_{\infty} h^{\mathrm{U}} + \frac{\tau_{\mathrm{w}}}{2} h^{\mathrm{R}} + \frac{\tau_{\mathrm{w}}}{2} h^{\mathrm{T}}.
\end{align}
\end{subequations}
Here, $\phi^{\mathrm{U}}$, $\phi^{\mathrm{R}}$, and $\phi^{\mathrm{T}}$ solve the following three auxiliary problems U, R, and T, respectively, and $h$ collectively denotes any of the macroscopic quantities $\omega, u_i, \tau, P, P_{ij}$, and $Q_i$.

\begin{figure*}
\centering
\subfigure[]{\includegraphics[scale=0.8]{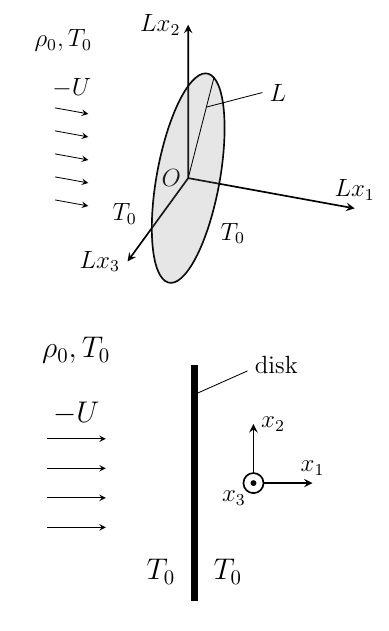}}
\subfigure[]{\includegraphics[scale=0.8]{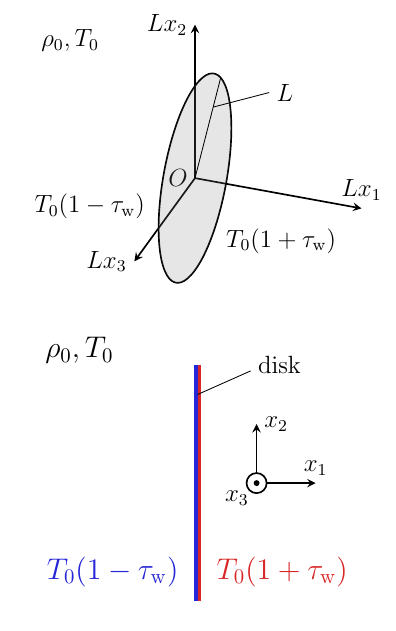}}
\subfigure[]{\includegraphics[scale=0.8]{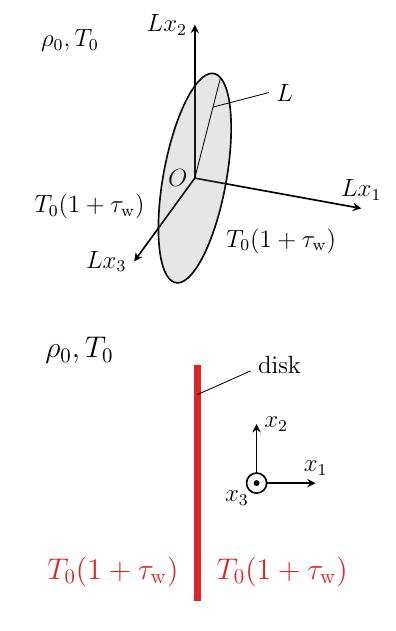}}
\caption{Schematics of the three auxiliary problems. 
(a) Problem U; (b) Problem R; (c) Problem T.
The bottom panels show the views from the positive  $x_3$-direction.}
\label{fig:decomp}
\end{figure*}

\noindent Problem U:
\begin{subequations}
\label{eq:prob_U}
\begin{align}
    \label{eq:LBGK_U}
    &\zeta_i \frac{\partial \phi^{\mathrm{U}}}{\partial x_i} = \frac{1}{\kappa} \mathcal{L}(\phi^{\mathrm{U}}), \\
\label{eq:bc_Sp_and_Sm_U}
    &\text{b.c. } \phi^{\mathrm{U}} = \sigma_{\mathrm{w}}^{\pm}(\phi^{\mathrm{U}}) \quad
     \text{for } \zeta_1 \gtrless 0 \quad (\bm{x} \in S^{\pm}), \\
\label{eq:bc_inf_U}
    &\text{b.c. } \phi^{\mathrm{U}} \to 2 \zeta_1 \quad (|\bm{x}| \to \infty).
\end{align}
\end{subequations}
Problem R:
\begin{subequations}
\label{eq:prob_R}
\begin{align}
    \label{eq:LBGK_R}
    &\zeta_i \frac{\partial \phi^{\mathrm{R}}}{\partial x_i} = \frac{1}{\kappa} \mathcal{L}(\phi^{\mathrm{R}}), \\
\label{eq:bc_Sp_and_Sm_R}
    &\text{b.c. } \phi^{\mathrm{R}} = \sigma_{\mathrm{w}}^{\pm}(\phi^{\mathrm{R}}) \pm (\zeta_j^2 - 2) \quad
     \text{for } \zeta_1 \gtrless 0 \quad (\bm{x} \in S^{\pm}), \\
\label{eq:bc_inf_R}
    &\text{b.c. } \phi^{\mathrm{R}} \to 0 \quad (|\bm{x}| \to \infty),
\end{align}
\end{subequations}
Problem T:
\begin{subequations}
\label{eq:prob_T}
\begin{align}
    \label{eq:LBGK_T}
    &\zeta_i \frac{\partial \phi^{\mathrm{T}}}{\partial x_i} = \frac{1}{\kappa} \mathcal{L}(\phi^{\mathrm{T}}), \\
\label{eq:bc_Sp_and_Sm_T}
    &\text{b.c. } \phi^{\mathrm{T}} = \sigma_{\mathrm{w}}^{\pm}(\phi^{\mathrm{T}}) + (\zeta_j^2 - 2) \quad
     \text{for } \zeta_1 \gtrless 0 \quad (\bm{x} \in S^{\pm}), \\
\label{eq:bc_inf_T}
    &\text{b.c. } \phi^{\mathrm{T}} \to 0 \quad (|\bm{x}| \to \infty).
\end{align}
\end{subequations}

The physical interpretations of these auxiliary problems are as follows. Problem U describes the flow of a rarefied gas past an infinitely thin circular disk whose surface is maintained at the same uniform temperature as the gas at infinity [see Fig.~\ref{fig:decomp}(a)]. Far from the disk, the gas is in an equilibrium state with density $\rho_0$, temperature $T_0$, and uniform flow velocity $(-U,0,0)$ ($|U| \ll (2RT_0)^{1/2}$). The corresponding dimensional VDF is given by $\rho_0 (2RT_0)^{-3/2} ( 1 + u_{\infty} \phi^{\mathrm{U}}) E$. In our previous study \cite{JoFM_TomitaEtAl_2026}, we obtained numerical solutions to Problem U over a wide range of Knudsen numbers. Note that $\phi^{\mathrm{U}}$ corresponds to $\phi/u_{\infty}$ in that study.

Problems R and T, on the other hand, describe the behavior of a rarefied gas around an infinitely thin circular disk whose surface temperature differs from the temperature of the otherwise undisturbed gas. In Problem R, $S^+$ and $S^-$ are maintained at two different uniform temperatures $T_0(1 + \tau_{\mathrm{w}})$ and $T_0(1 - \tau_{\mathrm{w}})$, respectively [see Fig.~\ref{fig:decomp}(b)], whereas in Problem T, both $S^+$ and $S^{-}$ are maintained at the same uniform temperature $T_0(1 + \tau_{\mathrm{w}})$ [see Fig.~\ref{fig:decomp}(c)], where $\tau_{\mathrm{w}} \ll 1$.
Far from the disk, the gas is in an equilibrium state at rest with density $\rho_0$ and temperature $T_0$. The corresponding dimensional VDFs are given by $\rho_0 (2RT_0)^{-3/2} ( 1 + \tau_{\mathrm{w}} \phi^{\mathrm{I}}) E$ ($\mathrm{I} = \mathrm{R}, \mathrm{T}$). Note that Problem R is antisymmetric, whereas Problem T is symmetric with respect to the plane of the disk.
Problem T can also be regarded as a formulation of thermal edge flow \cite{PoF_SoneYoshimoto_1997, Sone_2007,JoFM_TaguchiAoki_2012} around a circular disk.

In the present study, we consider the case in which $S^+$ is maintained at $T_0(1+\tau_{\mathrm{w}})$, while $S^-$ is maintained at $T_0$. More generally, suppose that $S^{+}$ and $S^{-}$ are maintained at $T_0 (1 + \alpha)$ and $T_0 (1 + \beta)$, respectively, where $|\alpha|$, $|\beta| \ll 1$, with all other conditions unchanged.
Under the linearization assumption, the corresponding solution is given by
\begin{align}
\label{eq:superposition_arbitrary}
    \phi = u_{\infty} \phi^{\mathrm{U}} + \frac{\alpha - \beta}{2} \phi^{\mathrm{R}} + \frac{\alpha + \beta}{2} \phi^{\mathrm{T}},
    \quad
    h = u_{\infty} h^{\mathrm{U}} + \frac{\alpha - \beta}{2} h^{\mathrm{R}} + \frac{\alpha + \beta}{2} h^{\mathrm{T}}.
\end{align}

We close this section by noting the spatial symmetries of Problems U, R, and T. It is readily confirmed from \eqref{eq:prob_U}--\eqref{eq:prob_T} that the solutions satisfy the following symmetry relations:
\begin{align}
\label{eq:symmetry}
    \phi^{\mathrm{I}}(x_1,x_2,x_3,\zeta_1,\zeta_2,\zeta_3) = \begin{cases}
        -\phi^{\mathrm{I}}(-x_1,x_2,x_3,-\zeta_1,\zeta_2,\zeta_3) &(\mathrm{I} = \mathrm{U}, \mathrm{R}), \\
        \phi^{\mathrm{I}}(-x_1,x_2,x_3,-\zeta_1,\zeta_2,\zeta_3) &(\mathrm{I} = \mathrm{T}).
    \end{cases}
\end{align}
Together with the definitions of macroscopic quantities in \eqref{eq:def_macro}, these relations lead to
\begin{align}
\label{eq:symmetry_macro_UR}
    h^{\mathrm{I}}(x_1,x_2,x_3) = \begin{cases}
        h^{\mathrm{I}}(-x_1,x_2,x_3) &(h = u_1, P_{12}, P_{13}, Q_1), \\
        -h^{\mathrm{I}}(-x_1,x_2,x_3) &(\text{other quantities}),
    \end{cases}
\end{align}
for $\mathrm{I} = \mathrm{U}, \mathrm{R}$, and
\begin{align}
\label{eq:symmetry_macro_T}
    h^{\mathrm{I}}(x_1,x_2,x_3) = \begin{cases}
        -h^{\mathrm{I}}(-x_1,x_2,x_3) &(h = u_1, P_{12}, P_{13}, Q_1), \\
        h^{\mathrm{I}}(-x_1,x_2,x_3) &(\text{other quantities}),
    \end{cases}
\end{align}
for $\mathrm{I} = \mathrm{T}$.
These symmetry relations allow us to restrict the analysis to the half-space $x_1 > 0$ by imposing \eqref{eq:symmetry} as a virtual boundary condition for $\zeta_1 > 0$ on $\{ (x_1, x_2, x_3) \mid x_1 = 0_+ \} \setminus S^{+}$.

\subsection{\label{subsec:drag_force}Force acting on the disk and terminal velocity}

We now return to the force acting on the disk.
In Problem B, the force exerted on the disk by the gas is parallel to the $x_1$-axis owing to axisymmetry. Denoting the dimensional force by $(F,0,0)$, $F$ is obtained by integrating the normal stress over the disk surface:
\begin{subequations}
\label{eq:force}
\begin{align}
\label{eq:force_acting_on_the_disk}
    &\frac{F}{p_0 L^2} = u_{\infty} f_D^{\mathrm{U}} + \frac{\tau_{\mathrm{w}}}{2} f_D^{\mathrm{R}}, \\
\label{eq:def_f_D}
    &f_D^{\mathrm{I}} := -4\pi \int_{0}^{1} P_{11}^{\mathrm{I}}(x_1=0_{+},r) \,r \,\mathrm{d}r 
    \quad (\mathrm{I} = \mathrm{U}, \mathrm{R}),
\end{align} 
\end{subequations}
where $r = \sqrt{x_2^2 + x_3^2}$, and use has been made of \eqref{eq:superposition_macro}, \eqref{eq:symmetry_macro_UR}, and \eqref{eq:symmetry_macro_T} for $h = P_{11}$.
Owing to the decomposition, the drag and radiometric contributions to the force are separated in \eqref{eq:force}; the corresponding terms in \eqref{eq:balance_in_ProbB} are $F_U = p_0 L^2 u_{\infty} f_D^{\mathrm{U}}$ and $F_R = p_0 L^2 \tau_{\mathrm{w}} f_D^{\mathrm{R}}/2$, respectively.
Note that $P_{11}^{\mathrm{T}}$, the normal stress in Problem T, does not contribute to the force, since its contributions from $S^{+}$ and $S^{-}$ cancel each other out.
Since $\kappa$ is the only parameter in the systems \eqref{eq:prob_U} and \eqref{eq:prob_R}, both $f_D^{\mathrm{U}}$ and $f_D^{\mathrm{R}}$ are functions of $\kappa$ alone. 

As explained following \eqref{eq:balance_in_ProbB}, the terminal velocity of the disk in Problem A is determined by requiring that the net force exerted by the gas, $F = F_{\mathrm{U}} + F_{\mathrm{R}}$, vanishes. Accordingly, defining the normalized terminal velocity by $u_{\infty *} :=- U_*(2RT_0)^{-1/2}$, we obtain
\begin{align}
\label{eq:u_infty_ast}
    u_{\infty *} = -\frac{f_{D}^{\mathrm{R}}}{2 f_{D}^{\mathrm{U}}} \tau_{\mathrm{w}}.
\end{align}
Thus, $u_{\infty *}$ is proportional to the temperature perturbation $\tau_{\mathrm{w}}$, with a proportionality coefficient that depends on $\kappa$ through the ratio $f_D^{\mathrm{R}}/f_D^{\mathrm{U}}$. Once Problems U and R have been solved, $u_{\infty *}$ can therefore be determined directly from \eqref{eq:u_infty_ast}, without solving the original system \eqref{eq:basic_system} for the unknown terminal velocity. Substituting $u_{\infty} = u_{\infty *}$ into \eqref{eq:superposition} yields
\begin{align}
\label{eq:phi_macro_ast}
    \phi_{*} := \frac{\tau_{\mathrm{w}}}{2} \left( -\frac{f_D^{\mathrm{R}}}{f_{D}^{\mathrm{U}}} \phi^{\mathrm{U}} + \phi^{\mathrm{R}} + \phi^{\mathrm{T}} \right), \quad
    h_{*} := \frac{\tau_{\mathrm{w}}}{2} \left( -\frac{f_D^{\mathrm{R}}}{f_{D}^{\mathrm{U}}} h^{\mathrm{U}} + h^{\mathrm{R}} + h^{\mathrm{T}} \right).
\end{align}
These expressions give the solution of Problem B for $U = U_*$ and hence the solution of Problem A. 

For later comparison, we also consider another special case of Problem B, namely $U = 0$. This corresponds to a disk fixed in an otherwise quiescent gas, a configuration investigated experimentally in \cite{PRE_SeldenEtAl_2009,JoFM_SeldenEtAl_2009}. Setting $u_{\infty} = 0$ in \eqref{eq:superposition} gives
\begin{align}
\label{eq:phi_macro_dagger}
    \phi_{\dagger} := \frac{\tau_{\mathrm{w}}}{2} \left( \phi^{\mathrm{R}} + \phi^{\mathrm{T}} \right), \quad
    h_{\dagger} := \frac{\tau_{\mathrm{w}}}{2} \left( h^{\mathrm{R}} + h^{\mathrm{T}} \right).
\end{align}

In summary, the radiometric force can be obtained from Problem R alone, while the terminal velocity $u_{\infty *}$ can be determined from Problems U and R.
The complete solution of Problem B can then be constructed from the solutions of Problems U, R, and T according to \eqref{eq:superposition_VDF}.
Since Problem U has already been solved in our previous study \cite{JoFM_TomitaEtAl_2026}, the remaining task is to solve Problems R and T numerically.
Note that $\kappa$ is the sole physical parameter in these auxiliary problems.

\section{\label{sec:numerical_method}Brief summary of the numerical method}

As previously mentioned, Problem U has already been solved numerically in our previous study \cite{JoFM_TomitaEtAl_2026}. Problems R and T are solved here using the same numerical method. In this section, we summarize only the key features of the method; the reader is referred to \cite{JoFM_TomitaEtAl_2026} for details.

The numerical scheme is an iterative method based on the integral form of the BGK equation, \eqref{eq:LBGK_U}, \eqref{eq:LBGK_R}, and \eqref{eq:LBGK_T} (see \cite{JoCP_TsujiAoki_2013, JoFM_TaguchiEtAl_2021} for applications to initial boundary-value problems). The integral form is derived using the method of characteristics. For a given $(\bm{x}, \bm{\zeta})$, the characteristic is the line through the point $\bm{x}$ parallel to $\bm{\zeta}$. Tracing the characteristic backward, it reaches either the disk or infinity, depending on $(\bm{x}, \bm{\zeta})$. These two possibilities divide the phase space $(\bm{x}, \bm{\zeta})$ into two regions separated by a surface across which the solution has a jump discontinuity. To account for these discontinuities numerically, the integral equation is discretized separately on the two sides of each discontinuity surface, so that the computational lattice does not cross the surface. This explicit treatment of the discontinuities is a key feature of the present numerical method.

Another difficulty arises from the infinite extent of the physical domain. In numerical computations, imposing the far-field conditions \eqref{eq:bc_inf_U}, \eqref{eq:bc_inf_R}, and \eqref{eq:bc_inf_T} directly is highly inefficient, since the deviations of the solutions from their far-field states decay only algebraically with the distance from the origin. To mitigate this difficulty, we employ far-field asymptotic forms derived with the aid of the asymptotic theory of kinetic equations \cite{Sone_2002, Sone_2007}. These asymptotic forms contain unknown constants, which are determined by numerically matching the asymptotic forms with the discretized solutions in the computational domain. For Problem U, the far-field asymptotic form is given in \cite{JoFM_TomitaEtAl_2026}. 
For Problem R, the same asymptotic form can be used after dropping the terms associated with the uniform flow (the first terms on the right-hand sides of (43b) and (A2) in \cite{JoFM_TomitaEtAl_2026}). Details of the numerical matching procedure, which is common to Problems U and R, are also given in \cite{JoFM_TomitaEtAl_2026}. 
For Problem T, on the other hand, we use a far-field asymptotic form, retained only to leading order, at which the temperature and density variations appear; the flow velocity and pressure enter only at higher orders [see \eqref{eq:asy_probT} in Appendix~\ref{sec:asy_probT}].

\section{\label{sec:numerical_results}Numerical results}

\subsection{\label{sec:drag_force_res}Radiometric force and terminal velocity}

\begin{figure*}
\centering
\subfigure[]{\includegraphics[scale=0.71]{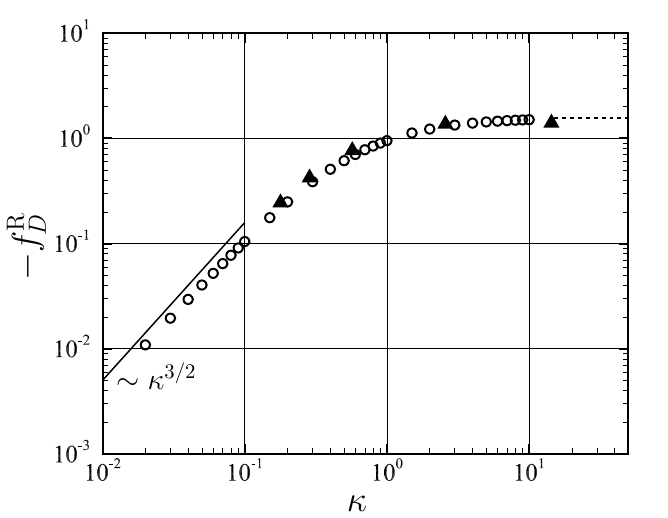}}
\subfigure[]{\includegraphics[scale=0.71]{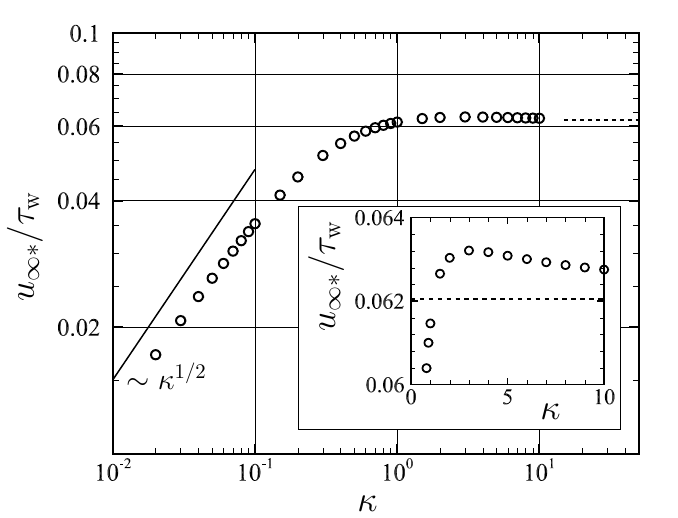}}
\caption{(a) Radiometric force $f_D^{\mathrm{R}}$ versus $\kappa$ (open circles). (b) Normalized terminal velocity $u_{\infty *}/\tau_{\mathrm{w}}$ versus $\kappa$, where $u_{\infty *}:=-U_*(2RT_0)^{-1/2}>0$. The solid lines indicate the asymptotic scalings as $\kappa \to 0$: (a) $\propto \kappa^{3/2}$; (b) $\propto \kappa^{1/2}$. The dashed lines indicate the limiting values as $\kappa \to \infty$: (a) $f_D^{\mathrm{R}} = -\pi/2$; (b) $u_{\infty *}/\tau_{\mathrm{w}} = \sqrt{\pi}/(4\pi+16)$.
In panel (a), the values deduced from the experimental results of Selden et al. \cite{JoFM_SeldenEtAl_2009} are also plotted by filled triangles.
The inset in panel (b) shows an enlarged view  around the maximum of $u_{\infty *}/\tau_{\mathrm{w}}$ on a linear scale.}
\label{fig:fD_R_and_tervel}
\end{figure*}

We first examine the dependence of $f_D^{\mathrm{R}}$ on $\kappa$. Figure~\ref{fig:fD_R_and_tervel}(a) shows $-f_D^{\mathrm{R}}$ as a function of $\kappa$ on double-logarithmic scales, with the numerical results indicated by open circles. (The numerical values of $-f_D^{\mathrm{R}}$ are summarized in Table~\ref{tab:sym_rel} in Sec.~\ref{sec:reciprocal_relation}.) The minus sign reflects the fact that the radiometric force acts in the negative $x_1$-direction in Fig.~\ref{fig:decomp}(b), i.e., from the hotter side toward the colder side. As seen from the figure, the magnitude of $f_D^{\mathrm{R}}$ increases monotonically with $\kappa$. In the continuum limit $\kappa \to 0$, the numerical results exhibit the $\kappa^{3/2}$ scaling, as indicated by the solid line. On the other hand, in the free-molecular limit $\kappa \to \infty$, $f_D^{\mathrm{R}}$ approaches a constant. This limiting value can be obtained analytically as $f_D^{\mathrm{R}} = -\pi/2 \approx -1.5708$, and is indicated by the dashed line.  

The experimental data of Selden et al. \cite{JoFM_SeldenEtAl_2009}, obtained for argon gas, are also shown by filled triangles. For comparison with the experiment, we approximately model argon atoms as hard spheres (HS) and convert the Knudsen number according to $\mathrm{Kn}_{\mathrm{BGK}} = 1.922284 \mathrm{Kn}_{\mathrm{HS}}$ so that the thermal conductivities of the BGK and HS models coincide \cite{Sone_2007}.
It should also be noted that, in most experiments \cite{PRL_PassianEtAl_2003, PRE_SeldenEtAl_2009, JoFM_SeldenEtAl_2009}, the ambient pressure $p_0$ is varied while the disk radius $L$ (or the size of the apparatus) and the temperature difference $\tau_{\mathrm{w}}$ are kept fixed. Under such conditions, the dimensional radiometric force $p_0 L^2 \tau_{\mathrm{w}} |f_D^{\mathrm{R}}| / 2$ attains a maximum at an intermediate value of $\kappa$, since $p_0 \propto \kappa^{-1}$, while $|f_D^{\mathrm{R}}| \sim \kappa^{3/2}$ as $\kappa \to 0$ and approaches a constant as $\kappa \to \infty$.

\begin{table}
\centering
\caption{Normalized terminal velocity $u_{\infty *}/\tau_{\mathrm{w}}$ versus $\kappa$, where $u_{\infty *}:=-U_*(2RT_0)^{-1/2}>0$.}
\label{tab:terminal_velocity}
\begin{tabular}{lclclclc} \hline\hline
    \multicolumn{1}{c}{$\kappa$} & $(u_{\infty *}/\tau_{\mathrm{w}}) \times 10^2$ &
    \multicolumn{1}{c}{$\kappa$} & $(u_{\infty *}/\tau_{\mathrm{w}}) \times 10^2$ &
    \multicolumn{1}{c}{$\kappa$} & $(u_{\infty *}/\tau_{\mathrm{w}}) \times 10^2$ &
    \multicolumn{1}{c}{$\kappa$} & $(u_{\infty *}/\tau_{\mathrm{w}}) \times 10^2$ \\ \hline
    $0.02$ & $1.7207$ & $0.1$ & $3.5240$ & $0.8$ & $6.0394$ & $6$ & $6.3004$ \\
    $0.03$ & $2.0737$ & $0.15$ & $4.1212$ & $0.9$ & $6.1001$ & $7$ & $6.2929$ \\
    $0.04$ & $2.3650$ & $0.2$ & $4.5520$ & $1$ & $6.1464$ & $8$ & $6.2862$ \\
    $0.05$ & $2.6152$ & $0.3$ & $5.1197$ & $1.5$ & $6.2654$ & $9$ & $6.2804$ \\
    $0.06$ & $2.8356$ & $0.4$ & $5.4664$ & $2$ & $6.3035$ & $10$ & $6.2753$ \\
    $0.07$ & $3.0327$ & $0.5$ & $5.6928$ & $3$ & $6.3210$ & $\infty$ & $6.2047$ \\
    $0.08$ & $3.2114$ & $0.6$ & $5.8480$ & $4$ & $6.3171$ & & \\
    $0.09$ & $3.3743$ & $0.7$ & $5.9585$ & $5$ & $6.3086$ & & \\ \hline\hline
\end{tabular}
\end{table}

Next, we present the terminal velocity of the disk discussed in Sec.~\ref{subsec:drag_force}, namely, $u_{\infty *}/\tau_{\mathrm{w}} = -f_D^{\mathrm{R}} / 2f_D^{\mathrm{U}}$ [see \eqref{eq:u_infty_ast}]. 
Here, we use the results for $f_D^{\mathrm{U}}$ obtained in \cite{JoFM_TomitaEtAl_2026}. 
Figure~\ref{fig:fD_R_and_tervel}(b) and Table~\ref{tab:terminal_velocity} summarize the values of $u_{\infty *}/\tau_{\mathrm{w}}$ for various $\kappa$. The figure shows its dependence on $\kappa$ on double-logarithmic scales. As $\kappa \to 0$, the terminal velocity exhibits the $\kappa^{1/2}$ scaling, since  $f_D^{\mathrm{R}} \propto \kappa^{3/2}$,
as shown above, whereas $f_D^{\mathrm{U}} \propto \kappa$ \cite{JoFM_TomitaEtAl_2026}. Indeed, the numerical results approach the solid line indicating the $\kappa^{1/2}$ scaling. This behavior is consistent with the previous numerical results for a two-dimensional configuration reported in \cite{JoCP_DechristeMieussens_2016}. In the free-molecular limit $\kappa \to \infty$, $u_{\infty *}/\tau_{\mathrm{w}}$ approaches $\sqrt{\pi}/(4\pi+16) \approx 0.062047$, as indicated by the dashed line in Fig.~\ref{fig:fD_R_and_tervel}(b). Notably, the terminal velocity is not monotonic in $\kappa$: it attains a maximum around $\kappa \approx 3$, slightly exceeding the corresponding value in the free-molecular limit. The inset in Fig.~\ref{fig:fD_R_and_tervel}(b) shows an enlarged view around the maximum on linear scales.

\subsection{\label{sec:flow_field}Flow field}

\begin{figure*}
\centering
\subfigure[]{\includegraphics[scale=0.57]{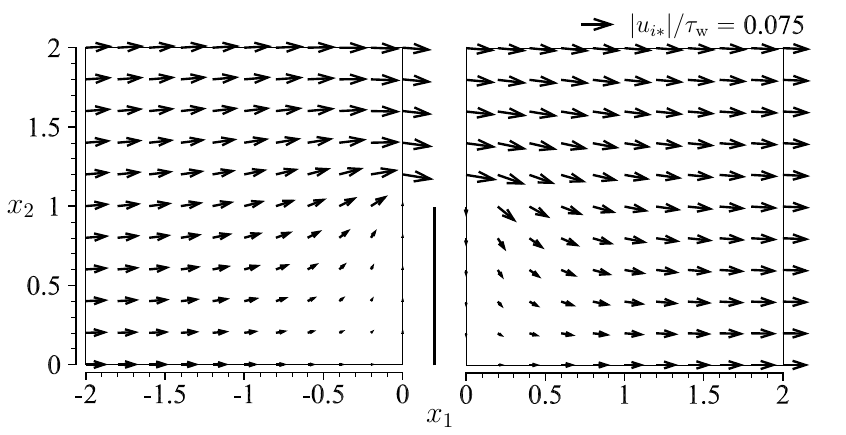}}
\subfigure[]{\includegraphics[scale=0.57]{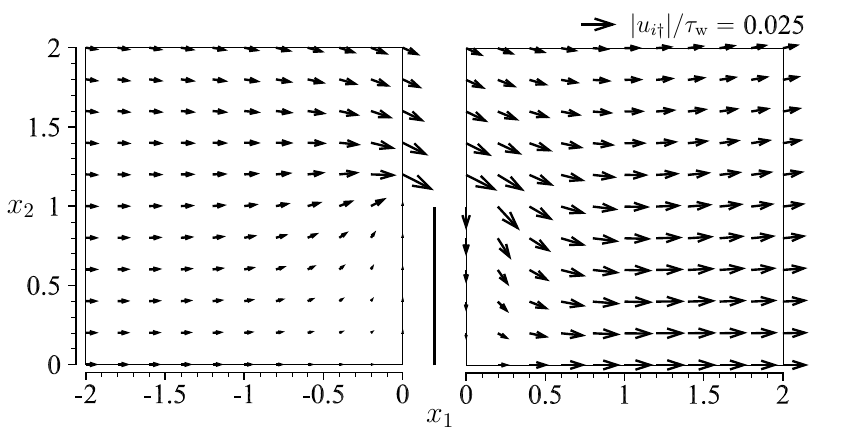}}
\subfigure[]{\includegraphics[scale=0.57]{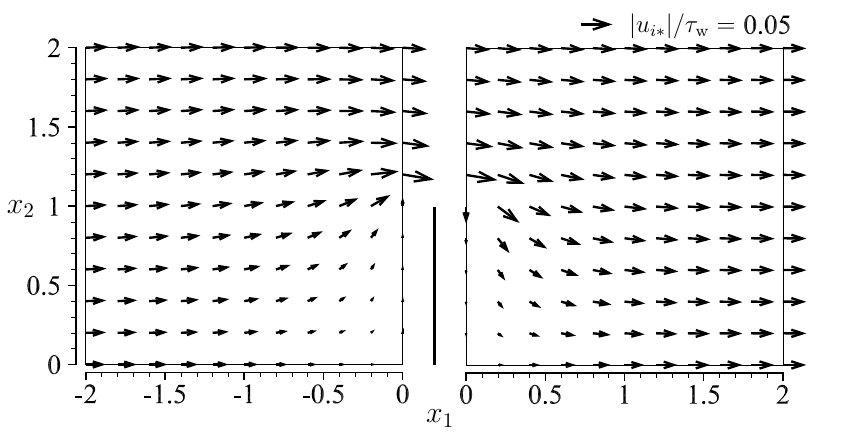}}
\subfigure[]{\includegraphics[scale=0.57]{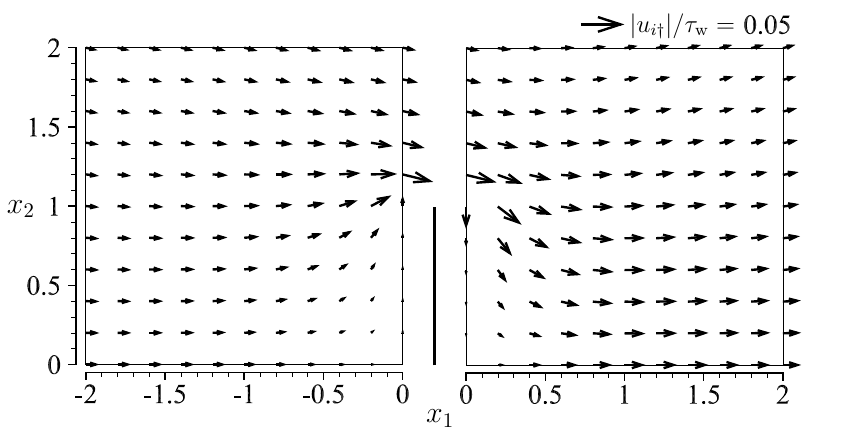}}
\caption{Flow fields in the region $-2 \leq x_1 \leq 2$, $0 \leq x_2 \leq 2$ in the plane $x_3=0$. Each vector represents the flow velocity at its starting point. The reference vector is displayed in the upper-right corner of each panel, and the solid line at the bottom center represents the disk. (a,b) $\kappa = 1$; (c,d) $\kappa = 0.1$. (a,c) $u_{i*}/\tau_{\mathrm{w}}$ for Problem A (or equivalently, Problem B with $U=U_*$); (b,d) $u_{i \dagger}/\tau_{\mathrm{w}}$ for Problem B with $U=0$.}
\label{fig:flow_field}
\end{figure*}
\begin{figure*}
\centering
\subfigure[]{\includegraphics[scale=0.6]{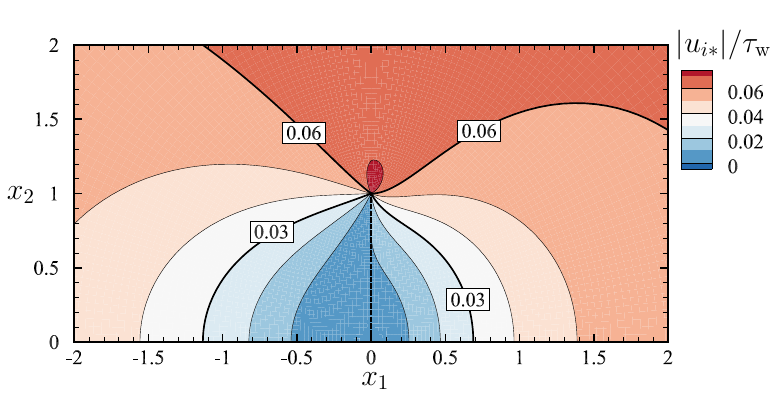}}
\subfigure[]{\includegraphics[scale=0.6]{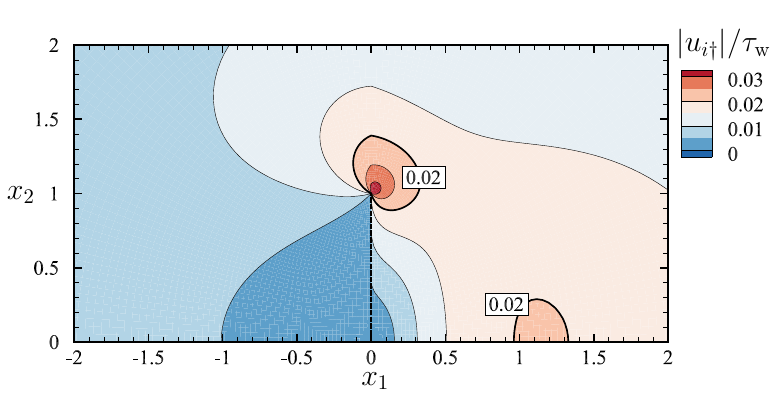}}
\subfigure[]{\includegraphics[scale=0.6]{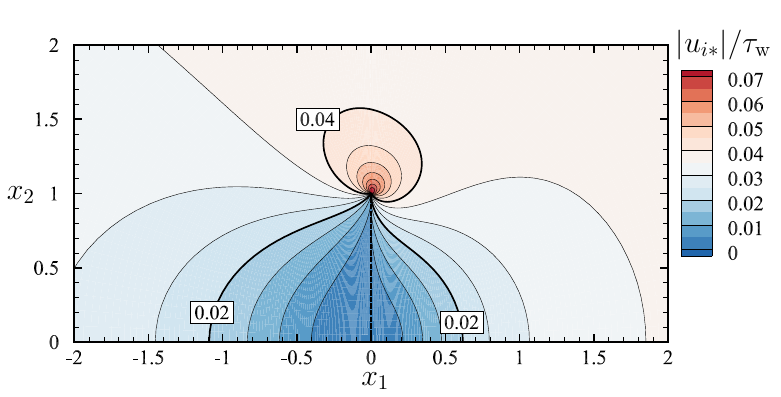}}
\subfigure[]{\includegraphics[scale=0.6]{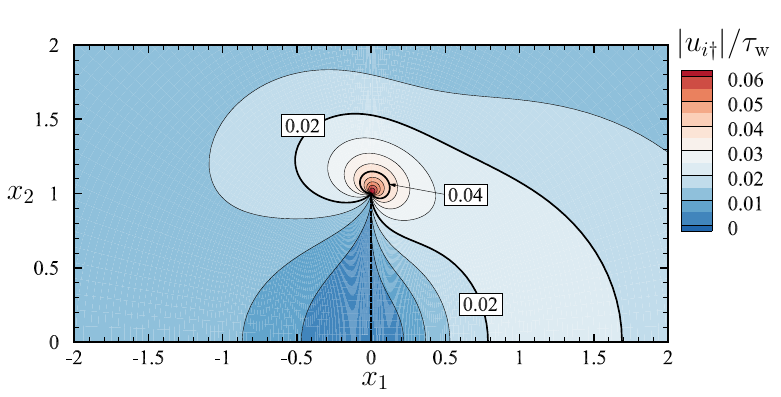}}
\caption{Spatial distributions of the flow speed corresponding to Fig.~\ref{fig:flow_field}. 
The solid line at $x_1=0$ and $0 \leq x_2 \leq 1$ represents the disk.
Some of the isolines are thickened and labeled for the sake of readability.}
\label{fig:flow_field_norm}
\end{figure*}

We next examine the flow field around the disk in two cases: the disk moving at the terminal velocity (Problem A, or equivalently, Problem B with $U = U_*$) and the stationary disk (Problem B with $U = 0$), as introduced in Sec.~\ref{subsec:drag_force}. 
Recall that subscripts $*$ and $\dagger$ are attached to the quantities in the former and latter cases, respectively [cf. \eqref{eq:phi_macro_ast} and \eqref{eq:phi_macro_dagger}].
In both cases, the results are presented in the reference frame fixed to the disk.

In Fig.~\ref{fig:flow_field}, we show the spatial distribution of $u_i$ for $\kappa=1$ [(a,b)] and $\kappa=0.1$ [(c,d)]. The left panels (a,c) show $u_{i*}$, while the right panels (b,d) show $u_{i \dagger}$. 
The velocity fields are shown in the plane $x_3=0$ over the region of $-2 \leq x_1 \leq 2$ and $0 \leq x_2 \leq 2$. Owing to the axisymmetry of the problem, any meridian plane gives the same profiles. To distinguish the behavior on the two sides of the disk, the regions $x_1 \gtrless 0$ are depicted separately. A reference vector is displayed at the upper-right corner of each panel. 
Note that $u_i$ is not uniquely defined at the disk edge $(x_1,x_2) = (0,1)$, since its limiting value depends on the direction of approach. In Fig.~\ref{fig:flow_field}, the value shown at the edge is the limiting value along the disk surface $S^{\pm}$, i.e., for $x_1 = 0_{\pm}$ and $x_2 \to 1-0$.
Figure~\ref{fig:flow_field_norm} shows contour maps of the flow speed $|u_i|[=(u_j^2)^{1/2}]$ under the same conditions as in Fig.~\ref{fig:flow_field}. These figures show that a stronger flow develops in a narrow region near the disk edge and becomes more localized at $\kappa = 0.1$. On the disk surface facing the positive $x_1$-direction, $S^+$, a non-zero tangential flow velocity is clearly observed in all cases. 

\begin{figure*}
\centering
\subfigure[]{\includegraphics[scale=0.6]{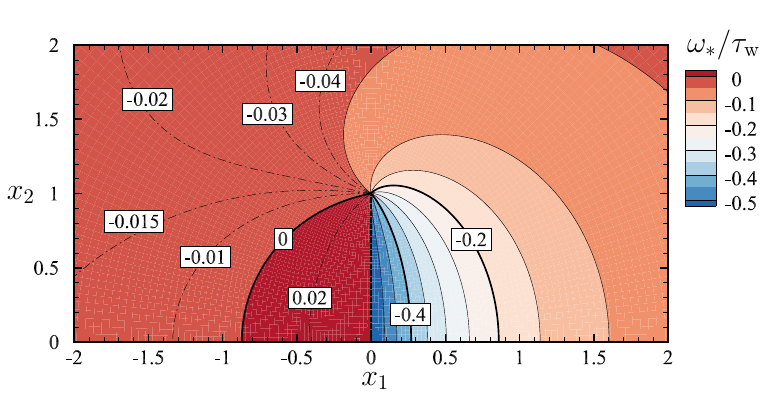}}
\subfigure[]{\includegraphics[scale=0.6]{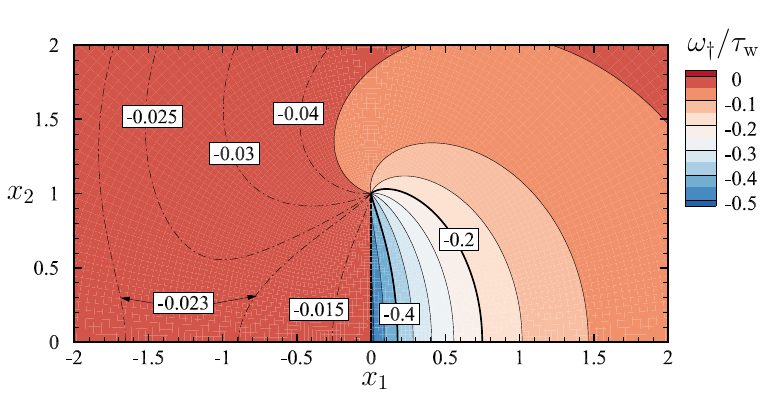}}
\subfigure[]{\includegraphics[scale=0.6]{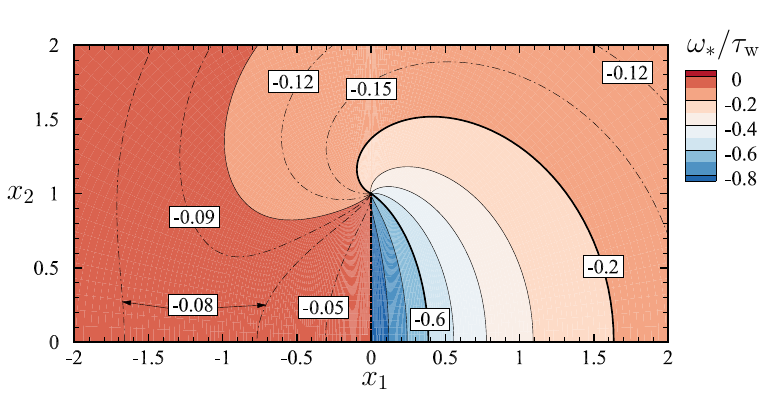}}
\subfigure[]{\includegraphics[scale=0.6]{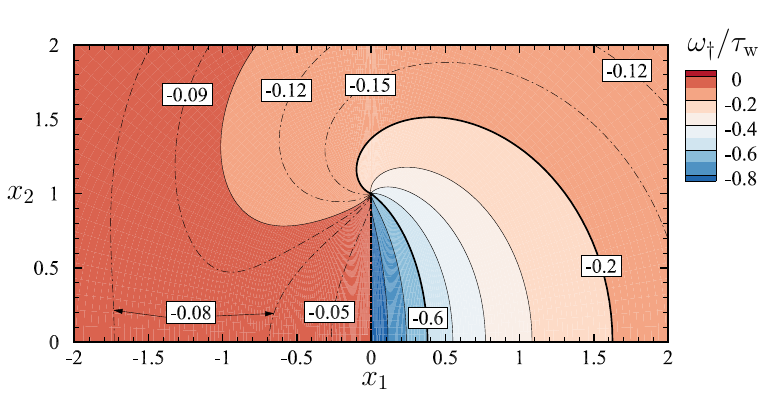}}
\caption{Density field. (a,b) $\kappa = 1$; (c,d) $\kappa = 0.1$. (a,c) $\omega_*/\tau_{\mathrm{w}}$; (b,d) $\omega_{\dagger}/\tau_{\mathrm{w}}$. Additional isolines are shown by dash-dotted lines. See also the caption of Fig.~\ref{fig:flow_field_norm}.}
\label{fig:density_field}
\end{figure*}
\begin{figure*}
\centering
\subfigure[]{\includegraphics[scale=0.6]{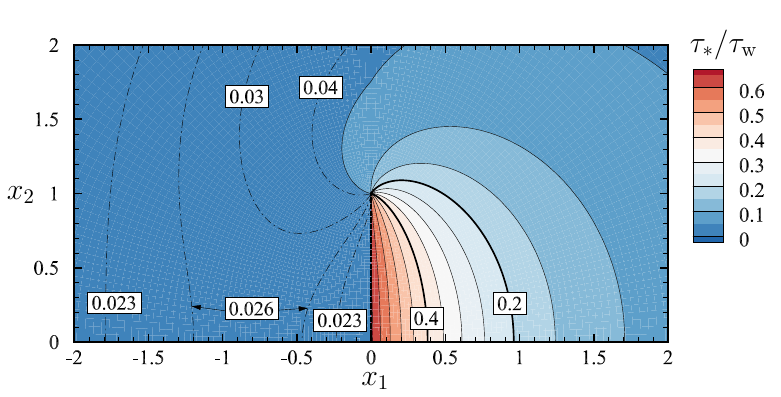}}
\subfigure[]{\includegraphics[scale=0.6]{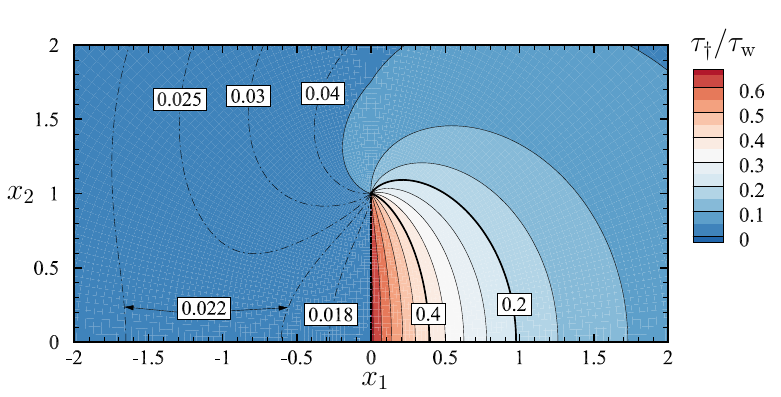}}
\subfigure[]{\includegraphics[scale=0.6]{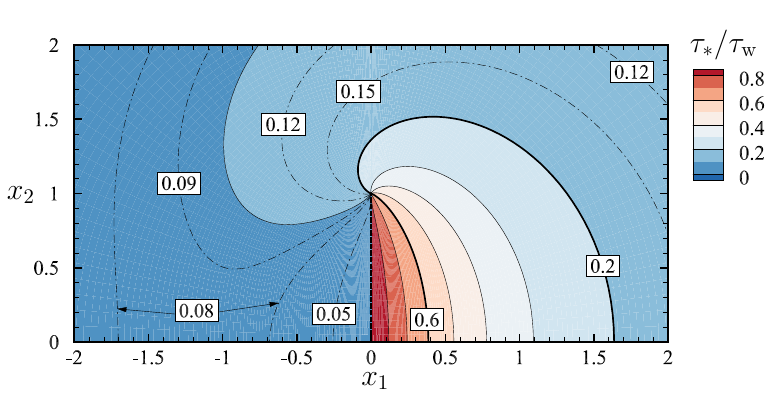}}
\subfigure[]{\includegraphics[scale=0.6]{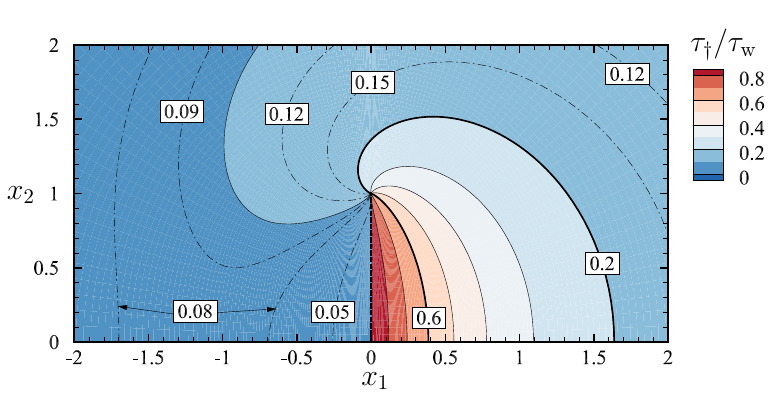}}
\caption{Temperature field. (a,b) $\kappa = 1$; (c,d) $\kappa = 0.1$. (a,c) $\tau_*/\tau_{\mathrm{w}}$; (b,d) $\tau_{\dagger}/\tau_{\mathrm{w}}$. Additional isolines are shown by dash-dotted lines. See also the caption of Fig.~\ref{fig:flow_field_norm}.}
\label{fig:temperature_field}
\end{figure*}
\begin{figure*}
\centering
\subfigure[]{\includegraphics[scale=0.6]{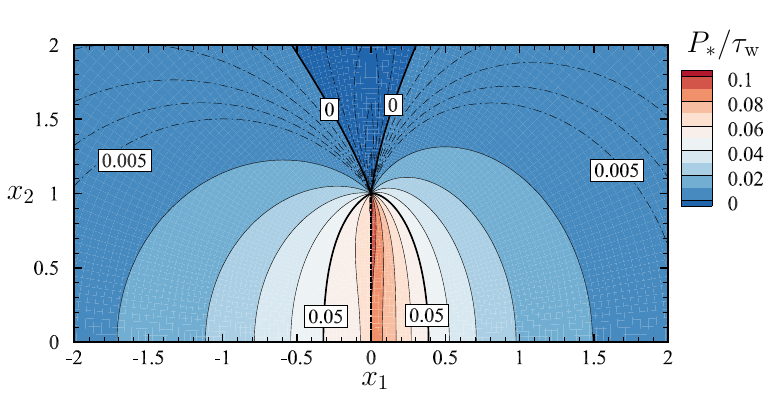}}
\subfigure[]{\includegraphics[scale=0.6]{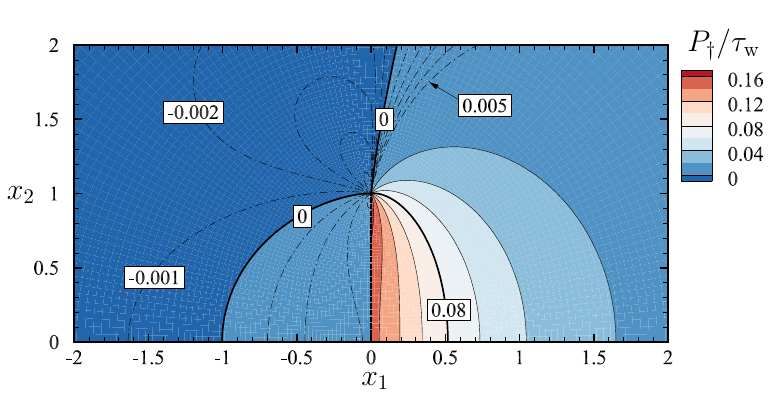}}
\subfigure[]{\includegraphics[scale=0.6]{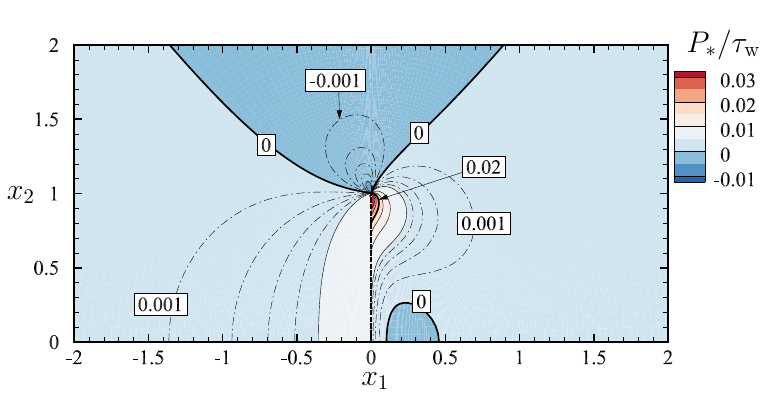}}
\subfigure[]{\includegraphics[scale=0.6]{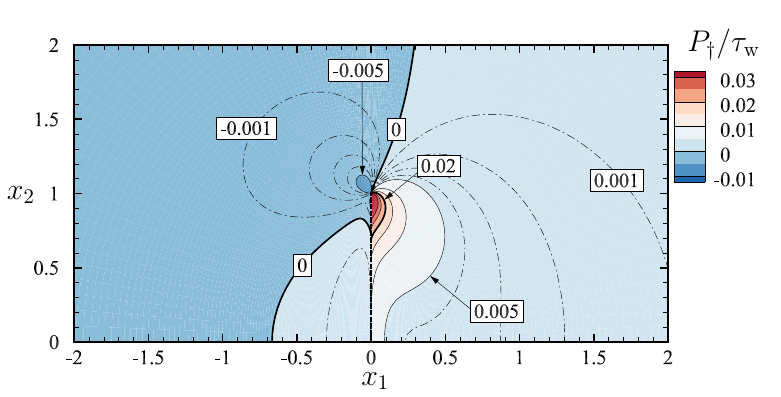}}
\caption{Pressure field. (a,b) $\kappa = 1$; (c,d) $\kappa = 0.1$. (a,c) $P_*/\tau_{\mathrm{w}}$; (b,d) $P_{\dagger}/\tau_{\mathrm{w}}$. Additional isolines at intervals of $0.001$ for $0 < |P/\tau_{\mathrm{w}}| \le 0.005$ are shown by dash-dotted lines. See also the caption of Fig.~\ref{fig:flow_field_norm}.}
\label{fig:pressure_field}
\end{figure*}

In Fig.~\ref{fig:density_field}--\ref{fig:pressure_field}, we show the spatial distributions of the density $\omega$, the temperature $\tau$, and the pressure $P$ in the region $-2 \leq x_1 \leq 2$, $0 \leq x_2 \leq 2$ in the plane $x_3=0$. The physical conditions in  each panel are the same as those in Fig.~\ref{fig:flow_field}. As shown in the figures, the density is lower in the region adjacent to the hotter side of the disk, whereas the temperature is higher. The overall distributions of the density and temperature are  similar for the freely moving disk [panels (a,c)] and the stationary disk [panels (b,d)], indicating that, under the prescribed surface-temperature condition considered here, the effect of the disk motion on these fields is relatively small. A temperature jump between the hotter surface of the disk and the adjacent gas is observed, and its magnitude is smaller in the case of $\kappa=0.1$. The pressure distributions, on the other hand, differ somewhat between the freely moving and stationary disks. Recall that the pressure perturbation is given by the sum of the density and the temperature, $P = \omega + \tau$ [see \eqref{eq:def_ppq}]. A pressure rise on the hotter side is observed in both cases. For the disk moving at the terminal velocity, a pressure rise also develops on the colder side owing to the disk motion, thereby reducing the pressure difference across the disk. This reflects the balance between the radiometric force and drag. A comparison between $\kappa = 1$ and $0.1$ reveals a significant difference: for $\kappa=0.1$, the high-pressure region on the hotter side is highly localized near the disk edge. This behavior, which illustrates the edge effect, is discussed further in the next section in terms of the normal stress $P_{11}$. We again note that the macroscopic quantities are not uniquely determined at the disk edge, and neither are their limiting values unless the direction of approach is specified.

\subsection{\label{sec:normal_stress}Normal stress on the disk}

\begin{figure*}
\centering
\subfigure[]{\includegraphics[scale=0.75]{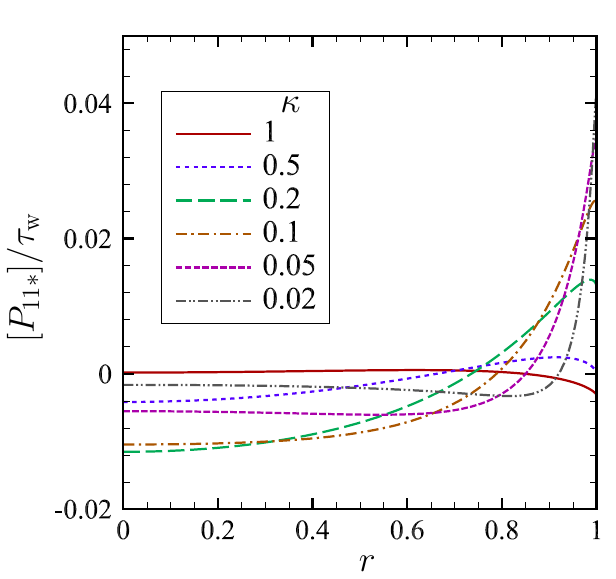}}
\subfigure[]{\includegraphics[scale=0.75]{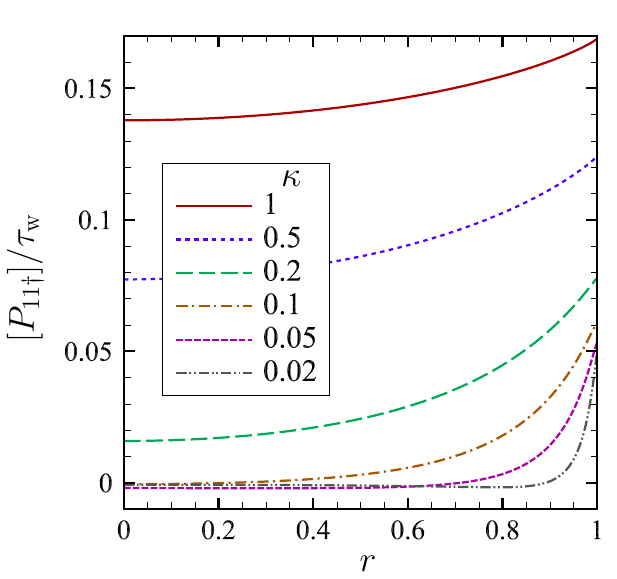}}
\caption{Distribution of the difference between the normal stresses on $S^{+}$ and $S^{-}$ for $\kappa = 1$, $0.5$, $0.2$, $0.1$, $0.05$, and $0.02$. (a) $[P_{11*}]/\tau_{\mathrm{w}}$, (b) $[P_{11\dagger}]/\tau_{\mathrm{w}}$.}
\label{fig:normal_stress}
\end{figure*}

In this section, we focus on the distribution of the normal stress $P_{11}$ on the disk surface, which, owing to the axisymmetry of the problem, can be expressed as a function of $r = \sqrt{x_2^2+x_3^2}$. Let us consider the difference in $P_{11}$ between the two sides of the disk, $S^{\pm}$, for Problem A (or equivalently, Problem B with $U = U_*$) and for Problem B with $U = 0$. We introduce a square bracket notation to denote this difference:
\begin{subequations}
\label{eq:def_diff_of_p11}
\begin{align}
\label{eq:def_diff_of_p11_ast}
    \left[ P_{11 *} \right] (r)
    &:= P_{11 *}(x_1=0_+, r) - P_{11 *}(x_1=0_-, r) \notag \\
    &= \tau_{\mathrm{w}} \left( -\frac{f_D^{\mathrm{R}}}{f_D^{\mathrm{U}}} P_{11}^{\mathrm{U}}(x_1=0_+, r) + P_{11}^{\mathrm{R}}(x_1=0_+, r) \right), \\
\label{eq:def_diff_of_p11_dagger}
    \left[ P_{11 \dagger} \right] (r)
    &:= P_{11 \dagger}(x_1=0_+, r) - P_{11 \dagger}(x_1=0_-, r)
     = \tau_{\mathrm{w}} P_{11}^{\mathrm{R}}(x_1=0_+, r).
\end{align} 
\end{subequations}
Here, we have used \eqref{eq:phi_macro_ast}, \eqref{eq:phi_macro_dagger}, and \eqref{eq:symmetry_macro_UR} in the last equalities.

We show the distributions of $\left[ P_{11*} \right]$ and $\left[ P_{11 \dagger} \right]$ in Fig.~\ref{fig:normal_stress}(a) and (b), respectively. More precisely, they are plotted as functions of $r$ for $\kappa = 1$, 0.5, 0.2, 0.1, 0.05, and 0.02. Positive values indicate that the net normal stress acts from the hotter side toward the colder side, whereas negative values indicate the opposite direction. 
For the freely moving disk [Fig.~\ref{fig:normal_stress}(a)], positive values of $\left[ P_{11 *} \right]$ correspond to a force locally acting in the direction of disk motion, whereas negative values correspond to a force acting against it. 
For small $\kappa$, the positive contribution that drives the disk is confined to a narrow region near the edge, while the normal-stress difference is negative over most of the disk surface. At the terminal velocity, the positive contribution to the force from the narrow region near the edge balances the negative contribution over the rest of the disk surface. 
For the stationary disk [Fig.~\ref{fig:normal_stress}(b)], on the other hand, $\left[P_{11 \dagger}\right]$ remains positive on the entire disk surface for all $\kappa$ shown, so that all parts of the disk contribute to the radiometric force. As $\kappa$ decreases, this contribution becomes increasingly concentrated near the edge.

As $\kappa$ increases, the normal-stress-difference profiles become flatter. In particular, the sign of $\left[P_{11*}\right]$ reverses from positive to negative near the edge and from negative to positive around the center. In the free-molecular limit $\kappa \to \infty$, the distributions become uniform: $[P_{11*}]/\tau_{\mathrm{w}} = 0$ and $[P_{11\dagger}]/\tau_{\mathrm{w}} = 1/4$ over the entire disk surface.
A similar edge-localized structure of the normal-stress difference was observed in a previous study of a two-dimensional periodic array of heated plates in a channel \cite{PRE_TaguchiAoki_2015}.
Note that the normal-stress difference in that reference was defined with
the opposite sign to that adopted in the present paper.

\begin{figure*}
\centering
\subfigure[]{\includegraphics[scale=0.75]{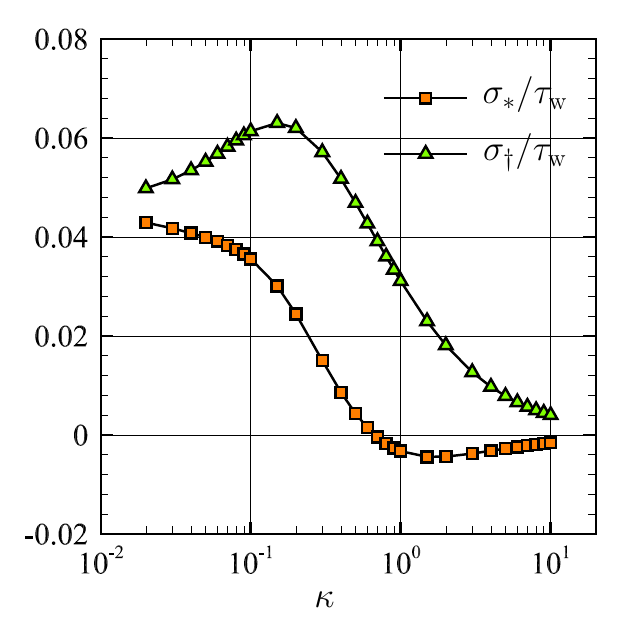}}
\subfigure[]{\includegraphics[scale=0.75]{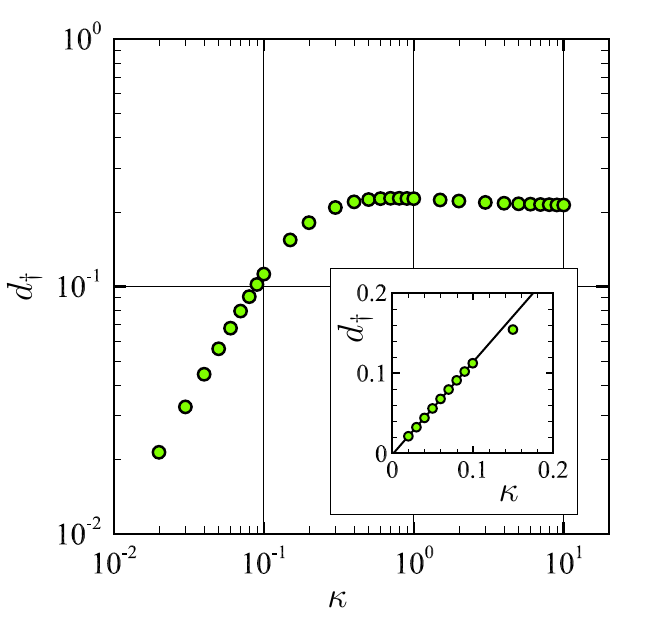}}
\caption{(a) $\sigma_*$ and $\sigma_{\dagger}$ versus $\kappa$. (b) $d_\dagger$ versus $\kappa$. The solid line in the inset represents the regression line \eqref{eq:LSM}, determined from the numerical values for $\kappa \leq 0.1$.}
\label{fig:peak_and_width}
\end{figure*}

Next, to characterize the edge-localized behavior of the normal-stress difference, we introduce
\begin{align}
\label{eq:def_sigma}
    \sigma_* = \lim_{r \to 1-0} [ P_{11*} ](r) - [P_{11*}](0), \quad
    \sigma_{\dagger} = \lim_{r \to 1-0} [ P_{11\dagger} ](r) - [P_{11\dagger}](0),
\end{align}
which measure the differences between the edge and center values of the normal-stress difference in Fig.~\ref{fig:normal_stress}(a) and (b), respectively.
Here, the values at the edge $r=1$ are defined as limiting values (see Sec.~\ref{sec:flow_field}).
Furthermore, for $U=0$, we introduce $d_\dagger$ by 
\begin{align}
\label{eq:def_d}
    [P_{11\dagger}](1-d_\dagger) = [P_{11\dagger}](0) + \frac{\sigma_{\dagger}}{2}.
\end{align}
Thus, $d_\dagger$ represents the distance from the edge at which $[P_{11\dagger}]$ reaches the midpoint between its center and edge values, and provides a measure of the width of the edge-localized region in Fig.~\ref{fig:normal_stress}(b).

In Fig.~\ref{fig:peak_and_width}, we show the values of $\sigma_*$, $\sigma_{\dagger}$, and $d_\dagger$ evaluated from the numerical results for various $\kappa$. 
From panel (a), we see that $\sigma_*$ attains a minimum around $\kappa \approx 1.5$, whereas $\sigma_\dagger$ attains a maximum around $\kappa \approx 0.15$. The change in sign of $\sigma_*$ around $\kappa \approx 0.7$ reflects the reversal of the relative magnitude of $[P_{11 *}]$ between the edge and center of the disk. As $\kappa \to \infty$, both $\sigma_{*}$ and $\sigma_{\dagger}$ vanish, since the normal-stress difference becomes uniform over the disk surface and the edge-localized behavior disappears. On the other hand, as $\kappa \to 0$, both quantities appear to approach nonzero limiting values. According to \cite{JoFM_TaguchiAoki_2012}, where radiometric flows were studied in a two-dimensional setting, such an edge-localized stress variation that remains finite in the continuum limit is associated with the thermal-stress effect. The present results suggest that a similar mechanism operates in the present three-dimensional configuration.
From panel (b), we see that $d_\dagger$ is proportional to $\kappa$ for small $\kappa$. In the inset, the numerical values are plotted on linear scales, together with the regression line obtained by the least-squares method using the data for $\kappa \leq 0.1$. The regression line is given by
\begin{align}
\label{eq:LSM}
    d_\dagger = a \kappa + b, \quad a = 1.1463, \quad b = -1.4169 \times 10^{-3}.
\end{align}
The coefficient of determination is 0.9997.
This result indicates that the characteristic width of the edge-localized normal-stress variation is of the order of the mean free path in the continuum limit. This finding is consistent with the assumption made by Einstein \cite{ZfP_Einstein_1924} and subsequently adopted in estimates of the radiometric force in more recent studies \cite{JoFM_SeldenEtAl_2009, PRF_BaierEtAl_2017}. 

\section{\label{sec:reciprocal_relation}Symmetric relation}

The radiometric effect can be regarded as a thermophoretic phenomenon, in which a body undergoes migration due to a non-uniform temperature distribution in the surrounding fluid or on the body surface. When both the speed of the body and the temperature variation are small, the reciprocal phenomenon, known as thermal polarization, can be deduced from Onsager's reciprocal relation \cite{Groot1962, JoNT_BakanovEtAl_1983, PASMaiA_Sharipov_1994a}. Namely, a non-uniform temperature field is induced by the motion of a body (or, equivalently, by a flow around it), even when the body has a uniform surface temperature distribution.
For a rarefied gas, Onsager's reciprocal relation can be derived from the symmetry properties of the linearized Boltzmann equation \cite{JoSP_Takata_2009, JoSP_Takata_2009a}. 
The symmetric relation provides an alternative way to compute certain global quantities \cite{JoSP_Takata_2009, JoFM_AokiEtAl_2014}.
In this section, by applying the symmetric relation to the present problem, we numerically validate the results for the radiometric force.

Among the three auxiliary problems introduced in Sec.~\ref{subsec:decomposition}, Problems R and U correspond to thermophoresis (the radiometric effect) and thermal polarization, respectively. The symmetric relation tells us that the radiometric force $f_{D}^{\mathrm{R}}$ in Problem R is equal to the integral of the heat flux over the disk surface in Problem U \cite{JoSP_Takata_2009}:
\begin{align}
\label{eq:sym_rel}
    f_D^{\mathrm{R}} = \mathcal{Q}^{\mathrm{U}}, \quad
    \mathcal{Q}^{\mathrm{U}} = -\int_{S^+ \cup S^{-}} g^{\mathrm{R}} Q_i^{\mathrm{U}} n_i \mathrm{d}S,
\end{align}
where $n_i$ is the unit normal vector on the disk surface $S^{+} \cup S^{-}$ pointing into the gas region, and $g^{\mathrm{R}}$ is the temperature distribution on the disk surface in Problem R, that is, $g^{\mathrm{R}} = 1$ on $S^+$ and $g^{\mathrm{R}} = -1$ on $S^-$.
Using \eqref{eq:symmetry_macro_UR} with $h = Q_1$ and $r = \sqrt{x_2^2 + x_3^2}$, together with axisymmetry, we obtain
\begin{align}
\label{eq:def_mathcalQU}
    \mathcal{Q}^{\mathrm{U}}
    = -\int_{S^+} Q_1^{\mathrm{U}} \mathrm{d}S - \int_{S^-} Q_{1}^{\mathrm{U}} \mathrm{d}S
    = -4\pi \int_{0}^{1} Q_1^{\mathrm{U}}(x_1=0_+,r) \,r \,\mathrm{d}r.
\end{align}
A step-by-step derivation of \eqref{eq:sym_rel} can also be found in Appendix A of \cite{JoFM_AokiEtAl_2014}. In that reference, an oblate spheroid with one hemispheroid slightly heated relative to the other is considered; the present disk geometry corresponds to the limiting case in which the aspect ratio tends to zero.

\begin{table}
\centering
\caption{Numerical values of $f_D^{\mathrm{R}}$ and $\mathcal{Q}^{\mathrm{U}}$ (i.e., the left- and right-hand sides of the first equation of \eqref{eq:sym_rel}) for various $\kappa$.}
\label{tab:sym_rel}
\begin{tabular}{lllclllclll} \hline\hline
\multicolumn{1}{c}{$\kappa$} & \multicolumn{1}{c}{$-f_{D}^{\mathrm{R}}$} & \multicolumn{1}{c}{$-\mathcal{Q}^{\mathrm{U}}$} & \quad\quad &
\multicolumn{1}{c}{$\kappa$} & \multicolumn{1}{c}{$-f_{D}^{\mathrm{R}}$} & \multicolumn{1}{c}{$-\mathcal{Q}^{\mathrm{U}}$} & \quad\quad &
\multicolumn{1}{c}{$\kappa$} & \multicolumn{1}{c}{$-f_{D}^{\mathrm{R}}$} & \multicolumn{1}{c}{$-\mathcal{Q}^{\mathrm{U}}$} \\ \hline
$0.02$ & $0.01088$ & $0.01094$ & & $0.2$ & $0.24872$ & $0.24870$ & & $2$ & $1.22474$ & $1.22465$ \\
$0.03$ & $0.01950$ & $0.01951$ & & $0.3$ & $0.38671$ & $0.38667$ & & $3$ & $1.33662$ & $1.33649$ \\
$0.04$ & $0.02942$ & $0.02939$ & & $0.4$ & $0.50827$ & $0.50820$ & & $4$ & $1.39572$ & $1.39557$ \\
$0.05$ & $0.04033$ & $0.04028$ & & $0.5$ & $0.61270$ & $0.61261$ & & $5$ & $1.43161$ & $1.43158$ \\
$0.06$ & $0.05204$ & $0.05199$ & & $0.6$ & $0.70192$ & $0.70179$ & & $6$ & $1.45576$ & $1.45573$ \\
$0.07$ & $0.06443$ & $0.06438$ & & $0.7$ & $0.77829$ & $0.77812$ & & $7$ & $1.47299$ & $1.47295$ \\
$0.08$ & $0.07734$ & $0.07729$ & & $0.8$ & $0.84398$ & $0.84377$ & & $8$ & $1.48581$ & $1.48580$ \\
$0.09$ & $0.09070$ & $0.09065$ & & $0.9$ & $0.90083$ & $0.90057$ & & $9$ & $1.49576$ & $1.49576$ \\
$0.1$ & $0.10441$ & $0.10436$ & & $1$ & $0.95034$ & $0.95005$ & & $10$ & $1.50369$ & $1.50368$ \\
$0.15$ & $0.17591$ & $0.17589$ & & $1.5$ & $1.12351$ & $1.12305$ & & & & \\ \hline\hline
\end{tabular}
\end{table}

We summarize the numerical values of $f_D^{\mathrm{R}}$ and $\mathcal{Q}^{\mathrm{U}}$ (i.e., the left- and right-hand sides of the first equation of \eqref{eq:sym_rel}) for various $\kappa$ in Table~\ref{tab:sym_rel}. These two values agree to at least three significant figures for all $\kappa$, supporting the reliability of the numerical results for Problems U and R. Finally, in light of the note in \cite{JoFM_AokiEtAl_2014}, we remark that a mathematical proof of the symmetric relation \eqref{eq:sym_rel} actually requires a rigorous treatment of the discontinuities in the VDFs $\phi^{\mathrm{U}}$ and $\phi^{\mathrm{R}}$, which was not explicitly addressed in \cite{JoSP_Takata_2009}. In this sense, the present result also provides numerical support for the validity of the symmetric relation in the presence of such discontinuities.

\section{\label{sec:conclusion}Concluding Remarks}

We numerically analyzed the radiometer effect on an infinitely thin circular disk immersed in a rarefied gas on the basis of the BGK equation and the diffuse reflection boundary condition. 
Under the assumption that the system can be linearized, the original problem was decomposed into three auxiliary problems, one of which describes the translational motion of the disk and has already been solved in our previous work \cite{JoFM_TomitaEtAl_2026}. In the present paper, we solved the remaining two problems and thereby obtained the thermally induced flow and the radiometric force. In the numerical computations, the discontinuities in the VDFs were carefully taken into account by employing the numerical scheme developed in \cite{JoFM_TomitaEtAl_2026}, which is based on the integral form of the BGK equation. 

The numerical results are summarized as follows. First, we presented numerical results for the radiometric force and the terminal velocity of the disk over a wide range of $\mathrm{Kn}$. In dimensionless form, the magnitude of the radiometric force increases monotonically with $\mathrm{Kn}$. It approaches a constant in the free-molecular limit $\mathrm{Kn} \to \infty$, whereas it scales as $\mathrm{Kn}^{3/2}$ in the continuum limit $\mathrm{Kn} \to 0$. The terminal velocity depends nonmonotonically on $\mathrm{Kn}$ and slightly exceeds its free-molecular-limit value at intermediate $\mathrm{Kn}$.
Second, we showed the spatial distributions of the flow velocity, density, temperature, and pressure around the disk for several typical values of $\mathrm{Kn}$ and illustrated the edge effect, the tangential flow along the disk surface, and the temperature jump. Third, we investigated the profiles of the normal-stress difference on the disk surface, with particular attention to its edge-localized behavior. In the continuum limit, the magnitude of the edge-localized variation appears to remain finite, whereas its characteristic width is of the order of the mean free path. 
Finally, using the symmetric relation, we validated the numerical results for the radiometric force.

We close this paper by commenting on some possible directions for future work. The radiometer effect can also be viewed as thermophoresis of a non-spherical particle in a gas. In this sense, the present numerical results may provide fundamental insight into the behavior of aerosol particles. Along this line, numerical computations based on more realistic physical models, in which the full Boltzmann equation is employed and/or the disk has a finite thickness, constitute a natural extension of the present work.
Another possible direction is to provide numerical support for further developments in asymptotic kinetic theory. For example, the scope of the generalized slip-flow theory \cite{Sone_2002, Sone_2007} is limited to problems with smooth boundary conditions. An attempt to remove this limitation and extend the theory to problems with non-smooth boundary conditions was made in \cite{JoFM_TaguchiTsuji_2020}, guided by prior numerical studies (e.g., \cite{PoF_AokiEtAl_2001}).
The numerical data obtained in the present paper and in \cite{JoFM_TomitaEtAl_2026} may serve as additional reference data for such theoretical developments.

\begin{acknowledgments}
This work was supported by JSPS KAKENHI Grant Numbers JP22K03924, JP22K18770, JP25K01156, and JP26K24580 and JST SPRING Grant Number JPMJSP2110. Numerical computations were carried out using the supercomputer of ACCMS, Kyoto University.
\end{acknowledgments}


\appendix

\section{\label{sec:asy_probT}Asymptotic forms in the far field}

In this appendix, we present the far-field asymptotic forms used in the numerical analysis in Sec.~\ref{sec:numerical_method}.
Suppose that the deviation of $\phi$ from its value at infinity, specified by \eqref{eq:bc_inf_U}, \eqref{eq:bc_inf_R}, and \eqref{eq:bc_inf_T}, decays algebraically with the distance from the origin, $\widetilde{r}=\sqrt{x_1^2 + x_2^2 + x_3^2}$. At a sufficiently large distance $\widetilde{r}_A$ such that $\mathrm{Kn}/\widetilde{r}_A \ll 1$, the behavior of $\phi$ on the spatial scale $\widetilde{r}_A$ is characterized by the small effective Knudsen number $\mathrm{Kn}_{\mathrm{eff}} = \mathrm{Kn}/\widetilde{r}_A$. According to the asymptotic theory of kinetic equations \cite{Sone_2002, Sone_2007}, the far-field behavior of the macroscopic quantities is governed by the Stokes set
\begin{subequations}
\label{eq:stokes_set_far_field}
\begin{gather}
\label{eq:Eq_cotinuum_far_field}
    \frac{\partial u_i}{\partial x_i} = 0, \\
\label{eq:Eq_Stokes_far_field}
    -\frac{\partial P}{\partial x_i} + \gamma_1 \kappa \frac{\partial^2 u_i}{\partial x_j^2} = 0, \\
\label{eq:Eq_Laplace_far_field}
    \frac{\partial^2 \tau}{\partial x_i^2} = 0, \\
\label{eq:Eq_state_far_field}
    \omega = P - \tau,
\end{gather} 
\end{subequations}
where $\gamma_1$ is the dimensionless viscosity coefficient; for the BGK model, $\gamma_1 = 1$.
Solutions of \eqref{eq:stokes_set_far_field} satisfying the far-field conditions appropriate to Problems U, R, and T provide the far-field asymptotic forms incorporated into the numerical computation.

Here, we present the far-field asymptotic form and its numerical implementation only for Problem T, since those for Problems U and R are described in \cite{JoFM_TomitaEtAl_2026}. (As noted in Sec.~\ref{sec:numerical_method}, the difference in the far-field treatment between Problems U and R is minor.) For Problem T, we use the following far-field asymptotic form:
\begin{subequations}
\label{eq:asy_probT}
\begin{align}
\label{eq:asy_omega_probT}
    &\omega^{\mathrm{T}} = -c_3 \mathrm{Arccot} (\sinh \xi), \\
\label{eq:asy_tau_probT}
    &\tau^{\mathrm{T}} = c_3 \mathrm{Arccot} (\sinh \xi), \\
\label{eq:asy_uP_probT}
    &u_i^{\mathrm{T}} = P^{\mathrm{T}} = 0,
\end{align}
\end{subequations}
where the oblate spheroidal coordinates $(\xi, \eta, \theta)$ are defined by $x_1 = \sinh \xi \cos \eta$, $x_2 = \cosh \xi \sin \eta \cos \theta$, and $x_3 = \cosh \xi \sin \eta \sin \theta$. The constant $c_3$ is a matching coefficient between the far-field asymptotic form and the numerical solution; its numerical determination is described below. 

The numerical implementation of \eqref{eq:asy_probT} is described below. According to the asymptotic theory \cite{Sone_2002, Sone_2007}, the corresponding far-field VDF is given by
\begin{align}
\label{eq:asy_phi_probT}
    \phi^{\mathrm{T}} \approx
    c_3 (\zeta^2 - \frac{5}{2}) \mathrm{Arccot} (\sinh \xi)
    + \kappa c_3 A(\zeta) \frac{\zeta_{\xi}}{(\sinh^2 \xi + \cos^2 \eta)^{1/2} \cosh \xi},
\end{align}
where $\zeta_{\xi}$ is the $\xi$-component of $\zeta_i$, and $A(\zeta)$ depends on the collision model; for the BGK model, $A(\zeta) = \zeta^2 - 5/2$. Expanding the right-hand side in inverse powers of $\chi = e^{\xi} \sim \widetilde{r}$ yields
\begin{align}
\label{eq:asy_phi_expand}
    \phi^{\mathrm{T}} = \phi^{\mathrm{T}}_{\mathrm{asy}} + O(\chi^{-3}), \quad
    \phi^{\mathrm{T}}_{\mathrm{asy}} = 2c_3 (\zeta^2 - \frac{5}{2}) \chi^{-1}.
\end{align}
To reduce the error caused by truncating the infinite domain, we solve for $\hat{\phi}^{\mathrm{T}} = \phi^{\mathrm{T}} - \phi^{\mathrm{T}}_{\mathrm{asy}}$ rather than directly for $\phi^{\mathrm{T}}$. Direct matching to determine $c_3$ would then require resolving the $O(\chi^{-3})$ terms, which is numerically difficult because these terms become comparable to numerical errors. We therefore determine $c_3$ using energy conservation. To this end, we define the total heat flux from the disk by
\begin{align}
\label{eq:def_mathcalQ}
    \mathcal{Q}^{\mathrm{T}} 
    := \int_{S^{+}} Q_1^{\mathrm{T}} \mathrm{d}S - \int_{S^{-}} Q_1^{\mathrm{T}} \mathrm{d}S
    = 4\pi \int_{0}^{1} Q_1^{\mathrm{T}}(x_1=0_+, r) \,r \,\mathrm{d}r,
\end{align}
where $\mathrm{d}S$ is the surface element and the second equality follows from axisymmetry and \eqref{eq:symmetry_macro_T} with $h=Q_1$. Energy conservation, together with \eqref{eq:asy_phi_probT} and the definition of the heat flux in \eqref{eq:def_ppq}, then gives 
\begin{align}
\label{eq:energy_conservation}
    c_3 = \frac{\mathcal{Q}^{\mathrm{T}}}{5\pi\gamma_2\kappa},
\end{align}
where $\gamma_2$ is the dimensionless thermal conductivity; for the BGK model, $\gamma_2 = 1$.
Since $\mathcal{Q}^{\mathrm{T}}$ is independent of $\chi$, \eqref{eq:energy_conservation} provides a numerically more robust
determination of $c_3$. At the end of each iteration for Problem T, \eqref{eq:energy_conservation} is used to update $c_3$ with an appropriate relaxation procedure.

\begin{table}[tbp]
\renewcommand{\arraystretch}{1.5}
\centering
\caption{Values of $\Delta_{f_D^{\mathrm{R}}}^{(\xi,\eta)}$, $\Delta_{h}^{(\xi,\eta)}$, $\Delta_{f_D^{\mathrm{R}}}^{(\alpha)}$, and $\Delta_h^{(\alpha)}$ ($h=\omega,u_1,u_r,\tau$; $\alpha = \zeta, \theta_{\zeta}, \varphi_{\zeta}$) for Problem R at $\kappa = 0.05$, $0.1$, and $1$.}
\label{tab:accuracy_check_probR}
\begin{tabular}{lccccc} \hline
 & $\Delta_{f_D^{\mathrm{R}}}^{(\xi,\eta)}$ & $\Delta_{\omega}^{(\xi,\eta)}$ & $\Delta_{u_1}^{(\xi,\eta)}$ & $\Delta_{u_r}^{(\xi,\eta)}$ & $\Delta_{\tau}^{(\xi,\eta)}$ \\ \hline
 $\kappa=1$ \,\,\, & \, $2.4 \times 10^{-5}$ \, & \, $7.9 \times 10^{-5}$ \, & \, $1.9 \times 10^{-4}$ \, & \, $6.8 \times 10^{-5}$ \, & \, $4.0 \times 10^{-5}$ \, \\
 $\kappa=0.1$ \,\,\, & $3.3 \times 10^{-4}$ & $8.3 \times 10^{-6}$ & $9.2 \times 10^{-5}$ & $3.5 \times 10^{-5}$ & $4.0 \times 10^{-6}$ \\
 $\kappa=0.05$ \,\,\, & $1.1 \times 10^{-3}$ & $1.3 \times 10^{-5}$ & $2.0 \times 10^{-4}$ & $7.3 \times 10^{-5}$ & $1.5 \times 10^{-5}$ \\ \hline
 & $\Delta_{f_D^{\mathrm{R}}}^{(\zeta)}$ & $\Delta_{\omega}^{(\zeta)}$ & $\Delta_{u_1}^{(\zeta)}$ & $\Delta_{u_r}^{(\zeta)}$ & $\Delta_{\tau}^{(\zeta)}$ \\ \hline
 $\kappa=1$ & $2.2 \times 10^{-6}$ & $9.6 \times 10^{-7}$ & $1.8 \times 10^{-6}$ & $1.1 \times 10^{-6}$ & $1.3 \times 10^{-6}$ \\
 $\kappa=0.1$ & $5.7 \times 10^{-6}$ & $2.5 \times 10^{-5}$ & $6.8 \times 10^{-6}$ & $6.1 \times 10^{-6}$ & $2.5 \times 10^{-5}$ \\
 $\kappa=0.05$ & $3.1 \times 10^{-5}$ & $9.1 \times 10^{-5}$ & $1.1 \times 10^{-5}$ & $1.1 \times 10^{-5}$ & $9.1 \times 10^{-5}$ \\ \hline
 & $\Delta_{f_D^{\mathrm{R}}}^{(\theta_{\zeta})}$ & $\Delta_{\omega}^{(\theta_{\zeta})}$ & $\Delta_{u_1}^{(\theta_{\zeta})}$ & $\Delta_{u_r}^{(\theta_{\zeta})}$ & $\Delta_{\tau}^{(\theta_{\zeta})}$ \\ \hline
 $\kappa=1$ & $8.7 \times 10^{-6}$ & $1.3 \times 10^{-4}$ & $2.1 \times 10^{-5}$ & $2.8 \times 10^{-4}$ & $1.6 \times 10^{-4}$ \\
 $\kappa=0.1$ & $5.2 \times 10^{-5}$ & $9.5 \times 10^{-5}$ & $9.4 \times 10^{-6}$ & $6.7 \times 10^{-4}$ & $1.3 \times 10^{-4}$ \\ 
 $\kappa=0.05$ & $6.9 \times 10^{-5}$ & $8.1 \times 10^{-5}$ & $7.9 \times 10^{-6}$ & $6.8 \times 10^{-4}$ & $1.2 \times 10^{-4}$ \\ \hline
 & $\Delta_{f_D^{\mathrm{R}}}^{(\varphi_{\zeta})}$ & $\Delta_{\omega}^{(\varphi_{\zeta})}$ & $\Delta_{u_1}^{(\varphi_{\zeta})}$ & $\Delta_{u_r}^{(\varphi_{\zeta})}$ & $\Delta_{\tau}^{(\varphi_{\zeta})}$ \\ \hline
 $\kappa=1$ & $3.3 \times 10^{-4}$ & $9.4 \times 10^{-5}$ & $5.8 \times 10^{-4}$ & $3.5 \times 10^{-4}$ & $2.6 \times 10^{-5}$ \\
 $\kappa=0.1$ & $5.2 \times 10^{-5}$ & $2.2 \times 10^{-5}$ & $5.4 \times 10^{-5}$ & $1.5 \times 10^{-5}$ & $2.6 \times 10^{-5}$ \\ 
 $\kappa=0.05$ & $4.1 \times 10^{-5}$ & $1.1 \times 10^{-4}$ & $3.4 \times 10^{-4}$ & $1.5 \times 10^{-5}$ & $1.2 \times 10^{-4}$ \\ \hline
\end{tabular}
\end{table}

\begin{table}[tbp]
\centering
\caption{Values of $\Delta_h^{(\xi,\eta)}$ and $\Delta_h^{(\alpha)}$ ($h=\omega,u_1,u_r,\tau$; $\alpha = \zeta, \theta_{\zeta}, \varphi_{\zeta}$) for Problem T at $\kappa = 0.1$ and $1$.}
\label{tab:accuracy_check_ProbT}
\renewcommand{\arraystretch}{1.5}
\begin{tabular}{lcccc} \hline
& $\Delta_{\omega}^{(\xi,\eta)}$ & $\Delta_{u_1}^{(\xi,\eta)}$ & $\Delta_{u_r}^{(\xi,\eta)}$ & $\Delta_{\tau}^{(\xi,\eta)}$ \\ \hline
$\kappa=1$ \,\,\, & \, $7.3 \times 10^{-5}$ \, & \, $7.1 \times 10^{-4}$ \, & \, $2.1 \times 10^{-4}$ \, & \, $3.7 \times 10^{-5}$ \, \\
$\kappa=0.1$ \,\,\, & $4.5 \times 10^{-6}$ & $8.4 \times 10^{-5}$ & $2.2 \times 10^{-5}$ & $4.5 \times 10^{-6}$ \\ \hline
& $\Delta_{\omega}^{(\zeta)}$ & $\Delta_{u_1}^{(\zeta)}$ & $\Delta_{u_r}^{(\zeta)}$ & $\Delta_{\tau}^{(\zeta)}$ \\ \hline
$\kappa=1$ & $6.0 \times 10^{-7}$ & $1.3 \times 10^{-6}$ & $8.0 \times 10^{-7}$ & $5.2 \times 10^{-7}$ \\
$\kappa=0.1$ & $2.6 \times 10^{-6}$ & $5.4 \times 10^{-6}$ & $6.3 \times 10^{-6}$ & $3.0 \times 10^{-6}$ \\ \hline
& $\Delta_{\omega}^{(\theta_{\zeta})}$ & $\Delta_{u_1}^{(\theta_{\zeta})}$ & $\Delta_{u_r}^{(\theta_{\zeta})}$ & $\Delta_{\tau}^{(\theta_{\zeta})}$ \\ \hline
$\kappa=1$ & $1.1 \times 10^{-4}$ & $6.2 \times 10^{-5}$ & $3.3 \times 10^{-4}$ & $1.4 \times 10^{-4}$ \\
$\kappa=0.1$ & $5.1 \times 10^{-5}$ & $4.0 \times 10^{-5}$ & $1.1 \times 10^{-3}$ & $7.2 \times 10^{-5}$ \\ \hline
& $\Delta_{\omega}^{(\varphi_{\zeta})}$ & $\Delta_{u_1}^{(\varphi_{\zeta})}$ & $\Delta_{u_r}^{(\varphi_{\zeta})}$ & $\Delta_{\tau}^{(\varphi_{\zeta})}$ \\ \hline
$\kappa=1$ & $8.8 \times 10^{-5}$ & $3.1 \times 10^{-4}$ & $1.5 \times 10^{-4}$ & $5.8 \times 10^{-5}$ \\
$\kappa=0.1$ & $4.3 \times 10^{-6}$ & $8.4 \times 10^{-5}$ & $2.5 \times 10^{-5}$ & $8.2 \times 10^{-6}$ \\ \hline
\end{tabular}
\end{table}

\section{\label{sec:accuracy}Accuracy of numerical solutions for Problems R and T}

In this appendix, we examine the dependence of the numerical results for Problems R and T on the lattice systems used in the computation. The numerical results presented in the main text for Problems U, R, and T were obtained using the same lattice systems. Details of these lattice systems, together with an accuracy test for Problem U, are given in Appendix B of \cite{JoFM_TomitaEtAl_2026}. Here, we apply the same procedure to Problems R and T. For completeness, we first summarize the notation for the lattice systems used in that reference.

The accuracy of the numerical results presented in this paper is assessed by comparison with the results obtained using finer lattices for typical values of $\kappa$. We refer to the lattice system used in the main numerical computations, detailed in Appendix B of \cite{JoFM_TomitaEtAl_2026}, as the ``standard setting.''
First, we simultaneously double the number of lattice points for the two spatial variables $(\xi, \eta)$. We define the relative variations in the radiometric force $f_D^{\mathrm{R}}$ and the macroscopic variables by
\begin{align}
\label{eq:def_Delta_xieta}
    \Delta_{f_D^{\mathrm{R}}}^{(\xi,\eta)} 
    = \frac{\left| f_D^{\mathrm{R}, (\xi,\eta)} - f_D^{\mathrm{R}, (\mathrm{sta})} \right|}
           {\left| f_D^{\mathrm{R}, (\mathrm{sta})} \right|}, \quad
    \Delta_h^{(\xi,\eta)} = \frac{\max_{i,j} \left| h_{2i,2j}^{(\xi,\eta)} - h_{i,j}^{(\mathrm{sta})} \right|}{\max_{i,j} \left| h_{i,j}^{(\mathrm{sta})} \right|},
\end{align}
where the superscripts $(\mathrm{sta})$ and $(\xi,\eta)$ on $f_D^{\mathrm{R}}$ and $h$ indicate results obtained with the standard and refined lattices, respectively, and $(i,j)$ denote the lattice-point indices for $(\xi, \eta)$. The refined lattice is indexed such that its points $(2i,2j)$ coincide with the points $(i,j)$ of the standard lattice. Thus, $\Delta_h^{(\xi,\eta)}$ measures the variation at the same spatial points. Next, we consider refinement of the molecular-velocity lattice, expressed in terms of the spherical coordinates $(\zeta, \theta_{\zeta}, \varphi_{\zeta})$ (see Sec.~2.3 of \cite{JoFM_TomitaEtAl_2026}). The number of lattice points for each variable is doubled separately, and the corresponding relative variations are defined by
\begin{align}
\label{eq:def_Delta_zeta}
    \Delta_{f_D^{\mathrm{R}}}^{(\alpha)} 
    = \frac{\left| f_D^{\mathrm{R}, (\alpha)} - f_D^{\mathrm{R}, (\mathrm{sta})} \right|}
           {\left| f_D^{\mathrm{R}, (\mathrm{sta})} \right|}, \quad
    \Delta_h^{(\alpha)} = \frac{\max_{i,j} \left| h_{i,j}^{(\alpha)} - h_{i,j}^{(\mathrm{sta})} \right|}{\max_{i,j} \left| h_{i,j}^{(\mathrm{sta})} \right|} \quad
    (\alpha = \zeta, \, \theta_{\zeta}, \, \varphi_{\zeta}),
\end{align}
where the superscript $(\alpha)$ indicates results obtained with the refined lattice in which the number of lattice points for $\alpha$ is doubled.

Table~\ref{tab:accuracy_check_probR} summarizes the values of $\Delta_{f_D^{\mathrm{R}}}^{(\xi,\eta)}$, $\Delta_{h}^{(\xi,\eta)}$, $\Delta_{f_D^{\mathrm{R}}}^{(\alpha)}$, and $\Delta_h^{(\alpha)}$ for Problem R, while 
Table~\ref{tab:accuracy_check_ProbT} summarizes $\Delta_{h}^{(\xi,\eta)}$ and $\Delta_h^{(\alpha)}$ for Problem T ($h=\omega, u_1, u_r, \tau$), where $u_r$ denotes the radial component
of the flow velocity. The evaluations were carried out at $\kappa=0.05$, $0.1$, and $1$ for Problem R and at $\kappa=0.1$ and $1$ for Problem T. As shown in the tables, the standard lattice provides sufficient accuracy for the representative cases examined here.

\bibliography{00bibTomita}

\end{document}